\documentclass[usenatbib]{mn2e}
\usepackage[letterpaper,margin=0.75in]{geometry}
\usepackage{natbibmnfix, graphicx, times, amsmath, epsfig, amssymb}
\usepackage{graphicx}
\usepackage{amsmath}
\usepackage{verbatim}
\usepackage{color}
\def\HH{${\rm {H_2}}\,\,$}  
\def\simgt{\lower.5ex\hbox{\gtsima}} 
\def\simlt{\lower.5ex\hbox{\ltsima}} 
\def\gtsima{$\; \buildrel > \over \sim \;$} 
\def\ltsima{$\; \buildrel < \over \sim \;$}
\def\Msun{M_\odot}

\def\HI{\hbox{H~$\scriptstyle\rm I\ $}}

\newcommand{\Tvir}{T_{\rm vir}}

\newcommand\lsim{\mathrel{\rlap{\lower4pt\hbox{\hskip1pt$\sim$}}
        \raise1pt\hbox{$<$}}}
\newcommand\gsim{\mathrel{\rlap{\lower4pt\hbox{\hskip1pt$\sim$}}
        \raise1pt\hbox{$>$}}}
\def\myputfigure#1#2#3#4#5%
{\vskip#5pt\makebox[0pt]{\hskip#2in
\includegraphics[width=#3\textwidth]{#1}}\vskip#4pt\hfill}

\newcommand{\apj}{ApJ}
\newcommand{\apjl}{ApJ}
\newcommand{\apjs}{ApJS}
\newcommand{\aap}{A\&A}
\newcommand{\aj}{AJ}
\newcommand{\mnras}{MNRAS}
\newcommand{\pasj}{PASJ}
\newcommand{\physrep}{Physics Reports}

\newcommand{\nat}{Nature}
\newcommand{\araa}{ARAA}

\begin{document}

\title[IMBH growth]
      {Simulating the growth of intermediate mass black holes}
\author[F. Pacucci \& A. Ferrara]
{Fabio Pacucci$^1$ \& 
Andrea Ferrara$^{1,2}$ \\
$^1$Scuola Normale Superiore, Piazza dei Cavalieri, 7  56126 Pisa, Italy \\
$^2$Kavli Institute for the Physics and Mathematics of the Universe (WPI), Todai Institutes for Advanced Study, the University of Tokyo \\
}
             
\date{submitted to MNRAS}

\maketitle
             
\begin{abstract}
Theoretical models predict that a population of Intermediate Mass Black Holes (IMBHs) of mass $M_\bullet \approx 10^{4-5} \, \mathrm{\Msun}$ might form at high ($z\gsim 10$) redshift by different processes. Such objects would represent the seeds out of which $z\simgt 6$ Super-Massive Black Holes (SMBHs) grow. We numerically investigate the radiation-hydrodynamic evolution governing the growth of such seeds via accretion of primordial gas within their parent dark matter halo of virial temperature $T_{vir} \sim 10^4 \, \mathrm{K}$. We find that the accretion onto a Direct Collapse Black Hole (DCBH) of initial mass $M_0=10^5 \, \mathrm{\Msun}$ occurs at an average rate $\dot{M}_{\bullet} \simeq 1.35 \, \dot{M}_{Edd} \simeq 0.1 \, \mathrm{M_{\odot} \, yr^{-1}}$, is intermittent (duty-cycle $\simlt 50$\%) and lasts $\approx 142 \, \mathrm{Myr}$; the system emits on average at super-Eddington luminosities, progressively becoming more luminous as the density of the inner mass shells, directly feeding the central object, increases. Finally, when $\approx 90\%$ of the gas mass has been accreted (in spite of an average super-Eddington emission) onto the black hole, whose final mass is $\sim 7 \times 10^6 \, \mathrm{\Msun}$, the remaining gas is ejected from the halo due to a powerful radiation burst releasing a peak luminosity $L_{peak}\sim 3\times 10^{45} \, \mathrm{erg \, s^{-1}}$. The IMBH is Compton-thick during most of the evolution, reaching a column density $N_H \sim 10^{25} \, \mathrm{cm^{-2}}$ in the late stages of the simulation. We briefly discuss the observational implications of the model.
\end{abstract}


\setcounter{footnote}{1}

\section{Introduction}
\label{sec:introduction}
In the standard $\Lambda$CDM cosmological model, the formation of the first stars occurred in the redshift range $30 \gtrsim z \gtrsim 20$ (\citealt{BL01, Miralda_E_2003, Bromm_Yoshida_2011, Stacy_2014, Hummel_2014}, for a recent and extensive review see also \citealt{Bromm_2013}) in mini-halos, i.e. dark matter structures with virial temperatures $\Tvir \simlt 10^4 \, \mathrm{K}$ and total  masses $M_h \simlt 10^8 \, \mathrm{M}_{\odot}$. The concurrent formation of the first black holes (\citealt{Bellovary_2011, Volonteri_Bellovary_2012, Agarwal_2013, Agarwal_2014}, see \citealt{Haiman_2013} for an updated review), as a final product of the evolution of the first massive stars, represents a second key event during the same cosmic epoch. The appearance of these classes of objects likely had a very strong impact on both the interstellar and the intergalactic medium, due to their radiative and mechanical feedback \citep{Ricotti_2011_parametric, Petri_2012, Ricotti_2012_DC, Jeon_2012, Tanaka_2012, Maiolino_2012, Ricotti_2013_luminosity, Nakauchi_2014}. 

The process of cooling and fragmenting the gas plays a role of key importance in this cosmic epoch.
Mini-halos primarily cool their metal-free gas through molecular hydrogen line emission. Under intense irradiation in the Lyman-Werner band (LW, $E_{\gamma}=11.2-13.6 \, \mathrm{eV}$), \HH is photo-dissociated via the two-step Solomon process (see the original study by \citealt{Draine_1996}), so that cooling (i.e. the formation of stars and stellar mass black holes) is quenched \citep{Visbal_2014}. The UV specific intensity in the LW band, $J_\nu= J_{21}^* \times 10^{-21}$ erg s$^{-1}$cm$^{-2}$Hz$^{-1}$\,sr$^{-1}$, required for quenching is somewhat uncertain, but lies in the range $J_{21}^*=0.1-1$ (see, e.g., \citealt{Machacek_2001} and \citealt{Fialkov_2013}).
When instead primordial, atomic-cooling halos ($T_{\rm vir}>10^4 \, \mathrm{K}$) are exposed to a LW flux of even higher intensity, $J_{\nu} > J_{\nu}^{\bullet}$ (\citealt{Loeb_Rasio_1994,Eisenstein_Loeb_1995,Begelman_2006,Lodato_Natarajan_2006, Regan_2009, Shang_2010,Johnson_2012, Agarwal_2012, Latif_2013}) the destruction of \HH molecules allows a rapid, isothermal collapse, initially driven by \HI Ly$\alpha$ line cooling, later replaced by two-photon emission. The precise value of $J_{\nu}^{\bullet}$ depends on several factors, but there is a general consensus that it should fall in the range $30 < J_{21}^{\bullet} < 1000$, depending on the spectrum of the sources \citep{Sugimura_2014}.

Several theoretical works (\citealt{Bromm_Loeb_2003, Begelman_2006, Volonteri_2008, Shang_2010, Johnson_2012}) have shown that the result of this collapse is the formation of a Direct Collapse Black Hole (DCBH) of mass $M_\bullet \approx 10^{4-5} \, \mathrm{\Msun}$, likely passing through the intermediate super-massive protostar phase, either directly collapsing into a compact object due to general relativistic instability (\citealt{Shibata_Shapiro_2002,Montero_2012}) or at the end of a very rapid evolution if the super-massive star reaches the Zero-Age Main Sequence (ZAMS, for a more detailed discussion see \citealt{Ferrara_2014}).  Once formed, the subsequent accretion of the remaining gas from the parent halo leads to a further growth of the DCBH into a fully-fledged Intermediate Mass Black Hole (IMBH) of mass $M_\bullet \approx 10^{5-6} \, \mathrm{\Msun}$.
This scenario is confirmed by cosmological simulations, as those presented by \cite{Latif_2013}, who have shown that under the previous conditions (atomic-cooling halos irradiated by $J_{\nu}>J_{\nu}^{\bullet}$) very strong accretion flows up to $\approx 1 \, \mathrm{M_{\odot}} \, \mathrm{yr^{-1}}$ may take place. As calculations by \cite{Hosokawa_2013} and \cite{Schleicher_2013} suggest that the effects of radiative feedback from the accreting protostar is weak, due to its cool ($\approx 6000$ K) photosphere, it has been possible to safely extend the simulations to longer time scales \citep{Latif_2013b}, up to the formation of a $\sim 10^5 \, \mathrm{M_{\odot}}$ central condensation\footnote{Even higher masses can result if magnetic fields, suppressing fragmentation, are included \citep{Latif_2014}.}.

In our work, we follow the accretion history onto a newly formed DCBH, with special emphasis on the dynamical and radiative properties of the inner parts of the parent halo, providing most of the accretion material to the central black hole. In particular, we aim at clarifying a number of aspects, including (i) the accretion time scale; (ii) the time-dependent emitted luminosity; (iii) the final outcome of the accretion phase, with a progressive depletion of the gas or a final outburst; (iv) the fraction of the halo gas accreted by the black hole and that ejected in the outskirts by radiative feedback.

Previous works have already attempted to describe the accretion process onto black holes of different sizes. \cite{Sakashita_1974} employed the method of similarity solutions to describe the time evolution of the accretion process in the optically thin regime (i.e. $\tau \lsim 1$, where $\tau$ is the optical depth). \cite{Tamazawa_1974}, instead, used a full radiative transfer approach to describe the process in the optically thick regime (i.e. $\tau \gsim 1$) assuming steady state, i.e. without any explicit time dependence in the accretion flow. In contrast with these previous works, we aim at a full time-dependent description in the optically thick regime, similarly to the somewhat idealized approach used by \citep{Ricotti_2011_parametric, Ricotti_2012_DC, Ricotti_2013_luminosity, Johnson_2013}, but extending these studies in several aspects and including a more complete physical description of the accretion process. Recent works (\citealt{Alexander_2014, Volonteri_2014,Madau_2014}) have proposed the occurrence of brief, recurring but strongly super-critical accretion episodes (with rates even $50-100$ times larger than the Eddington limit) to explain the rapid black hole mass build-up at high redshifts. An early phase of stable super-critical quasi-spherical accretion in the BHs was also proposed in \cite{Volonteri_2005}. Such large accretion rates may be sustainable in the so-called ``slim disk" solution (\citealt{Begelman_1982,Paczynski_1982,Mineshige_2000,Sadowski_2009,Sadowski_2011}), an energy-advecting, optically thick accretion flow that generalize the standard thin disk model \citep{Shakura_Sunyaev_1976}. In these radiatively inefficient models the radiation is trapped into the high-density gas and is advected inward by the accretion flow: as a consequence, the luminosity is only mildly dependent on the accretion rate and very large flows are sustainable. These works, while investigating the accretion flow at much smaller spatial scales, offer an interesting perspective in the discussion of the implications of our efforts, as detailed in Sec. \ref{sec:disc_concl}.

The present work is the first necessary step towards a precise prediction of the observable properties of IMBHs, whose existence has remained so far in the realm of theoretical, albeit physically plausible, speculations. This effort seems particularly timely given the advent of powerful telescopes as the James Webb Space Telescope (JWST), becoming available in the next few years, and high-$z$ ultra-deep surveys like the HST Frontier Fields \citep{Coe_2014}. In practice, we aim at determining the Spectral Energy Distribution of  IMBHs in order to build diagnostic tools able to uniquely identify these sources among the other high-$z$ ones. If successful, this strategy would not only represent a breakthrough in the study of the first luminous objects in the Universe, but it would also shed some light on the puzzles provided by the interpretation of the formation of SMBHs (see e.g. \citealt{Fan_2001, Mortlock_2011,Petri_2012}) and the excess recently found in the power spectrum of the cosmic Near Infrared Background fluctuations (for an overview, see \citealt{Yue_2013}).
These issues will be discussed in more details in Sec. \ref{sec:disc_concl}.

The outline of this paper is as follows. In $\S 2$ we describe the physics and equations of the radiation-hydrodynamic problem we aim at solving, along with the numerical implementation and initial conditions. In $\S 3$ we present the results of our simulations, providing a full picture of the accretion and feedback processes. Finally, in $\S 4$ we provide some discussion and the conclusions. The Appendix contains some more technical aspects of the simulations.
Throughout, we adopt recent Planck cosmological parameters \citep{Planck_Parameters_2013}: $(\Omega_m, \Omega_{\Lambda}, \Omega_b, h, n_s, \sigma_8 )= (0.32, 0.68, 0.05, 0.67, 0.96, 0.83)$.

\section{Physical and numerical implementation}
\label{sec:methods}
The present study is based on a series of radiation-hydrodynamic simulations. Our code is designed to execute a fully consistent treatment of uni-dimensional spherically-symmetric hydrodynamic (HD) equations and a simplified version of Radiative Transfer (RT) equations. 

The simulated spatial region is large enough to allow us to neglect deviations from spherical symmetry, e.g. the presence of an accretion disk which may form at much smaller scales. As detailed in \cite{deSouza_2013}, the primordial halos which generated the DCBHs rotate very slowly, with a mean value of the spin parameter $\lambda= Jc/(GM_h^2)=0.0184$, where $J$ is the angular momentum of the halo with mass $M_h$ and $c$ is the speed of light. Under these conditions, deviations from the spherical symmetry become important at the centrifugal radius $R_c =\lambda^2 G M_h^2/(c^2M_{\bullet}) \sim 10^{-6} \, \mathrm{pc} \sim 100 \, R_{\rm S}$, with $R_{\rm S}$ denoting the Schwarzschild radius: this value is $\sim 10^5$ times smaller than the internal boundary of our simulations. Our resolution is designed to resolve the Bondi radius (or gravitational radius, see \citealt{Bondi_1952}):
\begin{equation}
R_B = \frac{2GM_{\bullet}}{c_{s(\infty)}^2} = 1.5 \, \mathrm{pc} = 5\times10^{-4} \, R_{\rm vir} = 10^{8} \, R_{\rm S}
\end{equation}
where $R_{vir}$ is the virial radius of the halo and $c_{s(\infty)}$ is the sound speed at large distances from the accretion boundary, defined as:
\begin{equation}
c_{s(\infty)} = \sqrt{\gamma  \frac{p(R_B)}{\rho(R_B)}} \sim 15 \, \mathrm{km \, s^{-1}}
\end{equation}
Here, $\gamma = 5/3$ is the ratio of specific heats, $p(R_B)$ and $\rho(R_B)$ are the gas pressure and mass density at the Bondi radius, respectively.
The Bondi radius largely varies during our simulations, but, for clarity reasons, all distances in the plots are expressed in terms of its initial value $R_B(t=t_0)$.
Interestingly, even Adaptive Mesh Refinement cosmological simulations cannot resolve this spatial radius (see, e.g. \citealt{Pallottini_2013} where the maximum resolution is $\sim 5 \, \mathrm{kpc}$) and therefore they have to resort to some kind of sub-grid prescriptions for the black hole growth. In this context, the usual methodology to deal with black holes is to suppose that they irradiate with luminosities $L \approx L_{Edd}$ where:
\begin{equation}
L_{Edd} = 3.2 \times 10^4 \, \left(\frac{M_{\bullet}}{\rm M_{\odot}}\right) \, {\rm L_{\odot}} 
\label{eq:L_edd}
\end{equation}
is the Eddington luminosity. The domain of our simulations spans approximately from $0.1 \, R_B$ to $2 \, R_B$. This spatial dynamic range is designed to encompass with the highest possible resolution the spatial region of interest for the full simulation, i.e. from the radius of gravitational influence ($\sim R_B$) down to the smallest radius ($\sim 0.2 \, R_B$) reached by the propagating density wave (see Sec. \ref{sec:results}).
This spatial range has been successfully tested to verify the convergence of the main quantities derived in this work.
The natural time scale of the problem is given by the free-fall time at the Bondi radius:
\begin{equation}
t_{ff} \simeq \frac{1}{\sqrt{G\rho(R_B)}} \simeq 10^5 \, \mathrm{yr};
\label{tff}
\end{equation}
Here $\rho(R_B) \sim 3 \times 10^{-19} \, \mathrm{g \, cm^{-3}}$ is the mass density at the Bondi radius at the beginning of the simulations.
The time scale for the full RT simulation is about  $\sim 10^3 \, t_{ff}$, due to the presence of radiation pressure which slows down the collapse.

In the following subsections we describe the physics included in the simulations, separating the HD and the RT parts for clarity reasons. Some more technical aspects (boundary conditions, two-stream approximation, heating and cooling terms and photon diffusion) are deferred to the Appendix.

\subsection{Hydrodynamics}
We solve the standard system of ideal, non-relativistic  Euler's equations, i.e. neglecting viscosity, thermal conduction and magnetic fields, for a primordial (H-He) composition gas, spherically accreting onto a central DCBH, supposed at rest; the angular momentum of the gas with respect to the central object is zero. 

The code evolves in time the following system of conservation laws for mass, momentum and energy, solving for the radial component:
\begin{equation}
\frac{\partial \rho}{\partial t} + \nabla \cdot (\rho \mathbf{v}) = 0
\label{mass}
\end{equation}
\begin{equation}
\frac{\partial \mathbf{q}}{\partial t}  + \nabla \cdot (\mathbf{q} \times \mathbf{v})= -(\gamma -1)\nabla E -\nabla p_{rad} +\mathbf{g}\rho
\label{momentum}
\end{equation}
\begin{equation}
\frac{\partial E}{\partial t}  + \nabla \cdot (E \mathbf{v})= -(\gamma - 1) E \nabla \cdot \mathbf{v} + (H - C)
\label{energy}
\end{equation}
where $\rho$ is the gas mass density, $\mathbf{v}$ is the gas velocity (taken to be positive in the outward direction), $\mathbf{q}=\rho \mathbf{v}$ is the momentum density and $E$ is the energy density. Moreover $p_{rad}$ is the additional radiation pressure, $\mathbf{g}(r)$ is the gravitational field generated by the central black hole and $(H-C)$ are the heating and cooling terms. In the following, we neglect the vector notation since we only consider the radial components of the previous quantities. The total energy density $E$ is given by the relation:
\begin{equation}
E = \rho \epsilon + \frac{\rho v^2}{2}
\end{equation}
where $\epsilon$ is the specific gas thermal energy:
\begin{equation}
\epsilon = \frac{1}{\gamma - 1} \frac{1}{\mu} RT
\end{equation}
Here, $T$ is the gas temperature, $R$ is the gas constant and $\mu$ is the mean molecular weight which, for a primordial gas with helium fraction $Y_p = 0.2477$ \citep{Peimbert_2007} and no metals $Z_p=0$, is equal to $\mu=1.15$.
The gas thermal pressure is given by the usual equation for ideal gases:
\begin{equation}
P_g = \frac{1}{\mu} \rho R T, 
\label{EOS}
\end{equation}
while the gravitational acceleration is 
\begin{equation}
g(r,t) = \frac{GM_{\bullet}(t)}{r^2}
\end{equation}
The value of the black hole mass $M_{\bullet}(t)$ changes with time, due to the accretion, with the following set of rules, where $\dot{M}_{\bullet}=4 \pi r^2 \rho |v|$:
\begin{equation}
  \begin{cases}
   \dot{M}_{\bullet}(t) \neq 0 \, \, \,  \Leftrightarrow \, \,\, v(r=r_{min},t)<0 \\
   M_{\bullet}(t) = M_0 + (1-\eta)\int_0^t  \! \dot{M}_{\bullet} \, \mathrm{d}t
  \end{cases}
\end{equation}
where $M_0 = 10^5 \, {\rm M_{\odot}} $ is the initial value for the mass we adopt for the DCBH, $\dot{M}_{\bullet}$ denotes the time derivative of $M_{\bullet}$ and $\eta$ is the efficiency factor for mass-energy conversion (see the RT subsection below for more details).

The system Eqs. \ref{mass} - \ref{energy} are solved with a Linearized Roe Riemann solver \citep{Roe_1981}, a method based on Roe's matrix characteristic decomposition, which offers superior quality in treating the diffusion in hydrodynamic quantities. The time-stepping algorithm is a classical Runge-Kutta third-order scheme.
The time step is computed from the well-known CFL condition:
 \begin{equation}
v \, \frac{\Delta t}{\Delta r} = C< C_{max}
\end{equation}
where $C$ is the dimensionless Courant number and $C_{max} =0.8$.

\subsection{Radiative Transfer}
In our simulations, all radiation-related quantities are integrated over frequencies: this treatment serves as a good approximation for the main radiative features. Our RT modeling builds upon the works by \cite{Tamazawa_1974} for what concerns the general theoretical framework; we also exploit the computationally effective scheme by \cite{Novak_2012}. In the following we first introduce the relevant RT equations and then discuss their numerical implementation.

In the usual notation, $J$ is the intensity of the radiation field, $S$ is the source function, $H$ ($K$) is the first (second) moment of intensity. All these quantities are functions of time and position.
The closure relation between the second moment of intensity $K$ and the intensity $J$ is given by the so-called Eddington factor ${\cal F}$ ({\citealt{Hummer_1971, Tamazawa_1974}), here defined as:
\begin{equation}
{\cal F} = \frac{K}{J} = \frac{1 + \tau / \tau_0}{1 + 3 \tau / \tau_0}
\end{equation}
where $\tau_0$ is the optical depth at which the Eddington factor becomes equal to $1/2$.

The source total luminosity $L$ and $H$ are related at any point of the spatial domain by $L = 16 \pi^2 r^2 H$. In turn, $L$ depends on the gas accretion rate $\dot{M}_{\bullet}$ onto the IMBH (for $v<0$):
\begin{equation}
L_{\bullet} = \eta c^2 \dot{M}_{\bullet} = \eta c^2 \left( 4 \pi r^2 \rho |v| \right)
\end{equation}
where we employ an efficiency factor $\eta = 0.1$ (see e.g. \citealt{Yu_2002, Madau_2014}). Generally speaking, the efficiency factor ranges from $\eta = 0.057$ for a Schwarzschild (i.e. non-rotating) black hole to $\eta=0.32$ for a maximally rotating object (see \citealt{Thorne_1974}). We then define $L_{\bullet}$ the luminosity evaluated at the innermost grid cell, located at $r=r_{min}$:
\begin{equation}
\begin{cases}
L_{\bullet} \equiv  \eta c^2 \left[ 4 \pi r_{min}^2 \rho(r_{min}) |v(r_{min})| \right] & \text{if } v(r_{min}) <0 \\
L_{\bullet} \equiv 0 & \text{if } v(r_{min}) \geq 0
\end{cases}
\label{L_BH_definition}
\end{equation}
In addition, we define $f_{Edd} \equiv \dot{M}_{\bullet}/\dot{M}_{Edd}$: the accretion is super-critical (super-Eddington) if $f_{Edd}>1$. Note that, only in the case of a fixed value of the efficiency factor $\eta$, $f_{Edd} \equiv \dot{M}_{\bullet}/\dot{M}_{Edd} = L/L_{Edd}$ holds.
The acceleration caused by radiation pressure is then:
\begin{equation}
a_{rad} = \frac{\kappa(\rho,T) L}{4 \pi r^2 c}
\label{a_rad}
\end{equation}
where $\kappa(\rho,T)$ is the total opacity of the gas: in our simulations the Thomson electron scattering, bound-free and $H^-$ opacity terms are included. The Thomson opacity is  calculated following the temperature-dependent prescription given in \cite{Begelman_2008}, namely:
\begin{equation}
\kappa_T(T) = 0.2 (1+\mu) \frac{1}{1+(T/T_{*})^{-\beta}} \, \mathrm{cm^2 \, g^{-1}},
\end{equation}
with $\beta=13$. Below $T\sim T_{*} = 8000 \, \mathrm{K}$, the opacity rapidly falls due to the decrease of the ionized fraction: as a consequence, also the effectiveness of the radiation pressure is quenched.
The bound-free opacity is computed through the Kramers' approximation for a metal-free gas ($Z=0$) (see e.g. \citealt{Introduction_Modern_Astrophysics}):
\begin{equation}
\kappa_{bf}(\rho,T) = 4\times10^{22} (1+X_p) \rho T^{-3.5} (1-\chi_{ion})
\end{equation}
where $X_p$ is the primordial hydrogen fraction.
The opacity due to the negative hydrogen ion $H^-$ is computed as (see again e.g. \citealt{Introduction_Modern_Astrophysics}):
\begin{equation}
\kappa_{H^-}(\rho,T) = 3.5\times10^{-27} \rho^{0.5} T^{7.7} (1-\chi_{ion})
\end{equation}
With a metal-free gas ($Z=0$), within the ranges of mass densities, temperatures and ionized fractions in our spatial domain, the dominant source of opacity is the Thomson one.
Our definition of the optical depth $\tau$ is:
\begin{equation}
\tau(R) \equiv - \int_{0}^{R} \kappa \rho \, \mathrm{d}r
\end{equation}
The quantity reported in the graphs is the optical depth for an external observer.

As we will see, due to feedback effects, accretion onto the central object occurs in an intermittent manner. It is then useful to introduce the duty-cycle, defined as the fraction of time spent accreting within a given time frame of duration $T_{tot}$:
\begin{equation}
{\cal D} \equiv \frac{\Delta t_{accr}}{T_{tot}}
\end{equation}
The instantaneous value of the duty-cycle is computed dividing the total integration time $T_{tot}$ in slices and computing the duty-cycle with respect to each slice.

The equations for steady and spherically-symmetric transfer of radiation have been derived, e.g. by \cite{Castor_1972} and \cite{Cassinelli_Castor_1973}. For our problem, it is appropriate to assume steady-state RT equations since the light-crossing time at the Bondi radius, $R_B/c \approx 5 \, \mathrm{yr}$, is negligible with respect to the free-fall time, see Eq. \ref{tff}. The full RT equations are:
\begin{align}
\nonumber &\frac{v}{c} \frac{d}{dr} \left( \frac{4 \pi J}{\rho} \right) + 4 \pi K \frac{v}{c} \frac{d}{dr} \left( \frac{1}{\rho} \right) - \frac{4 \pi}{\rho} \frac{v}{c} \left(\frac{3K - J}{r}\right) = \\
&=-\frac{1}{4 \pi \rho r^2} \frac{dL}{dr} -4 \pi \kappa(J-S)
\label{eq:full_RT_1}
\end{align}
\begin{align}
\frac{dK}{dr} + \left(\frac{3K - J}{r}\right) + \frac{v}{c}\frac{dH}{dr} -\frac{2}{r} \frac{v}{c} H -\frac{2}{\rho} \frac{v}{c} \frac{d\rho}{dr} H= -\rho \kappa H
\label{eq:full_RT_2}
\end{align} \\
The term $\kappa(J-S)$ handles the gas heating and cooling, as the $(H-C)$ terms in the energy equation, see Eq. \ref{energy}.
These equations are correct up to the first order in $\beta = v/c$ and are suitable for high-speed accretion flows, where $\beta$ is not negligible. In the previous equations, the transition from the optically thin to the optically thick regime is described by the density-dependent Eddington factor ${\cal F}$ (decreasing from $1$ to $1/3$ with increasing optical depth) and by the term $\sim \rho \kappa(\rho, T)$.

\cite{Novak_2012} presented several computationally-effective non-relativistic RT formulations which yield the correct behavior both in optically thin and optically thick regimes. These formulations are convenient because they allow to obtain accurate results with a lower computational complexity. We defer the interested reader to the full derivation in \cite{Novak_2012} and we write down only the final formulation.
\begin{equation}
\frac{dL}{dr} = 4 \pi r^2 (\dot{E} - 4 \pi \rho \kappa J)
\label{N1}
\end{equation}
\begin{equation}
\frac{dJ}{dr} = -\frac{2 J w}{r} - \frac{(3-2w) \rho \kappa L}{16 \pi^2 r^2}
\label{N2}
\end{equation}
In these expressions, $w$ is an interpolation parameter which controls the transition from optically thin to optically thick regimes and is a function of the position: it ranges from $w=0$ (for an isotropic radiation field, i.e. in the optically thick regime) to $w=1$ in the optically thin regime. Furthermore, $\dot{E}$ is the source term, playing the same role of $S$ in the relativistic treatment (see the Appendix for a more detailed description). Eq. \ref{N1} is derived from the full RT Eqs. \ref{eq:full_RT_1}-\ref{eq:full_RT_2} by neglecting the $v/c$ terms and using $\dot{E}$ instead of $S$. Eq. \ref{N2}, instead, is derived by using the specific values of the Eddington factor ${\cal F} \equiv K/J$ and the interpolation factor $w$ in each regime: (${\cal F}=1$, $w=1$) for the optically thin and (${\cal F}=1/3$, $w=0$) for the optically thick.

Finally, we need to specify the boundary conditions. As in the study of stellar interiors, the second order differential equation for $L(r)$, or the two first-order ODEs in $L(r)$ and $J(r)$ (Eqs. \ref{N1}-\ref{N2}), requires two boundary conditions, at the inner and outer boundaries.
The innermost cell of the grid radiates the luminosity produced by the accretion flow onto the black hole (see Eq. \ref{L_BH_definition}), so that $L(r_{min}) \equiv  L_{\bullet}$.
Far from the black hole, the luminosity is expected to resemble a point source because the scattering becomes negligible, so that:
\begin{equation}
J(r_{max}) \equiv \frac{L(r_{max})}{16 \pi^2 r_{max}^2}
\end{equation}
In order to solve the set of boundary-value differential Eqs. \ref{N1}-\ref{N2}, the so-called shooting method with Newton-Raphson iteration \citep{Numerical_Recipes} works well up to the optical depth of a few. Beyond this limit, the shooting method becomes unstable and it is necessary to switch to a more powerful relaxation method \citep{Numerical_Recipes}.
However, following the evolution of the system for physically relevant time scales ($\sim 100 \, \mathrm{Myr}$) requires such a large number of steps that even the relaxation method becomes computationally not viable. To overcome this problem, we follow the two-stream approximation method outlined in the Appendix B of \cite{Novak_2012} and sketched in our Appendix.

\begin{table*}
\begin{minipage}{170mm}
\begin{center}
\caption{Simulation parameters for the two test simulations (T1 and T2) and for the full one (FS) which includes a complete solution of the radiative transfer.}
\label{tab:outline}
\begin{tabular}{|c|c|c|c|}
\hline\hline
Parameter & T1 & T2 & FS\\
\hline
$r_{min}$ [pc] & $0.16 $    & $0.16$  & $0.16$  \\
$r_{max}$ [pc] & $3.2  $    &$3.2 $ &$3.2$ \\
Integration time [yr]  & $3.2 \times 10^{4}$    & $3.2 \times 10^{4} $ & $1.4 \times 10^{8} $\\
Radiation pressure & no & yes & yes \\
Energy equation & adiabatic & adiabatic & non-adiabatic  \\
\end{tabular}
\end{center}
\end{minipage}
\end{table*}

\subsection{Initial Conditions}
We model the spherically-symmetric gas accretion onto a seed black hole of initial mass $M_0=10^5 \, \mathrm{M}_{\odot}$, assumed at rest at the center of a dark matter halo of virial temperature $T_{vir} \sim 10^4 \, \mathrm{K}$ and total mass (dark matter and baryons) $M_h  = 6.2 \times 10^7 \, \mathrm{M}_{\odot}$ at redshift $z=10$. For $r \ll R_{vir}$ most of the mass is baryonic: for this reason we supposed that the gas mass contained in our computational domain ($r<2R_B \ll R_{vir}$) is $M_{gas} = (\Omega_b/\Omega_m) \, M_h = 9.6 \times 10^6 \, \mathrm{M_{\odot}}$ (similarly to what has been done in \citealt{Latif_2013}, see their Fig. 5). This assumption is reasonable since, during the formation process of the DCBH, most of the baryonic mass of the halo is expected to fall into its gravitational influence (the halo has no spin, then no centrifugal barrier is formed). 

We assume that the gas follows the density profile derived from the simulations in \cite{Latif_2013}, which is well approximated by the following functional form: 
\begin{equation}
\rho(r) = \frac{\rho_0}{1+(r/a)^2}
\end{equation}
where $a$ is the core radius, estimated from Fig. 1 of \cite{Latif_2013} as $a \sim 270 \, \mathrm{AU} \sim 1.3 \times 10^{-3} \, \mathrm{pc}$: this scale is too small to be resolved by our simulations.
Normalization to the gas mass content, $M_{gas}$, gives the central density value $\rho_{0} \sim 10^{-11} \, \mathrm{g\,cm^{-3}}$.
From the initial prescription for the density field and assuming an isothermal profile with $T= 10^4 \, \mathrm{K}$ (as in \citealt{Latif_2013} and leading to a very small value, $\sim 10^{-3}$, of the ionized fraction throughout the domain), the initial conditions for the pressure field are derived directly from the equation of state, Eq. \ref{EOS}. The initial conditions for the radial velocity are set to a very small, spatially constant value $v_0 <0$. After a very brief ($ \ll t_{ff}$) transient, the system adjusts to a velocity profile consistent with a rapid accretion across the inner boundary of the grid. In addition, the initial density profile is also rapidly modified within the Bondi radius, while outside the gravitational influence of the black hole it remains almost unaltered.

The spatial grid for all simulations extends from $r_{min} = 5.0\times 10^{17} \, \mathrm{cm} \approx 0.16 \, \mathrm{pc}$ to $r_{max} = 1.0\times 10^{19} \, \mathrm{cm} \approx 3.2 \, \mathrm{pc}$, with $3000$ logarithmically spaced bins. 
This range is optimized for our computational target, allowing to follow the entire displacement of the density wave and extending out to the Bondi radius.
In our work we do not have the sufficient spatial resolution to resolve the $\mathrm{H \, II}$ region produced by the accretion-generated radiation (but see e.g. \citealt{Milosavljevic_2009} where the $\mathrm{H \, II}$ region is resolved). A simple estimate of the extension of the Stromgren radius in our case yields a dimension of $\sim 0.03 \, \mathrm{pc}$, which is in very good accordance with the radius reported in \cite{Ricotti_2011_parametric}.

For further details about the simulations, e.g. the boundary conditions, see the Appendix.

\section{Results}
\label{sec:results}
We present our results starting from the two test simulations T1 (pure hydro) and T2 (hydro + radiation pressure) in order to highlight some physical key features of the problem. Next, we turn to the analysis of the full simulation, FS, that in addition includes the complete solution of the radiative transfer; the FS simulation contains our main findings. The simulation parameters used in the three runs are reported in Table \ref{tab:outline}.

The total integration time $T_{tot}$ for the runs T1 and T2 has been chosen in order to show the most important features of their radiation-hydrodynamic evolution, while for the FS simulation it corresponds to the total history of the system.

\subsection{Test simulations}
\subsubsection{T1: Pure hydro}
Simulation T1 is a purely adiabatic hydrodynamic simulation. The only two forces acting on the gas are produced by the gravitational field of the black hole and by the pressure gradient; therefore, it corresponds to the ``classical" Bondi solution \citep{Bondi_1952}, with the only exception consisting in the limited gas reservoir, which prevents a genuine steady state to take place. The spatial profiles predicted by T1 are shown in Fig. \ref{fig:HD_spatial}.

\begin{figure*}
	\includegraphics[angle=0,width=0.45\textwidth]{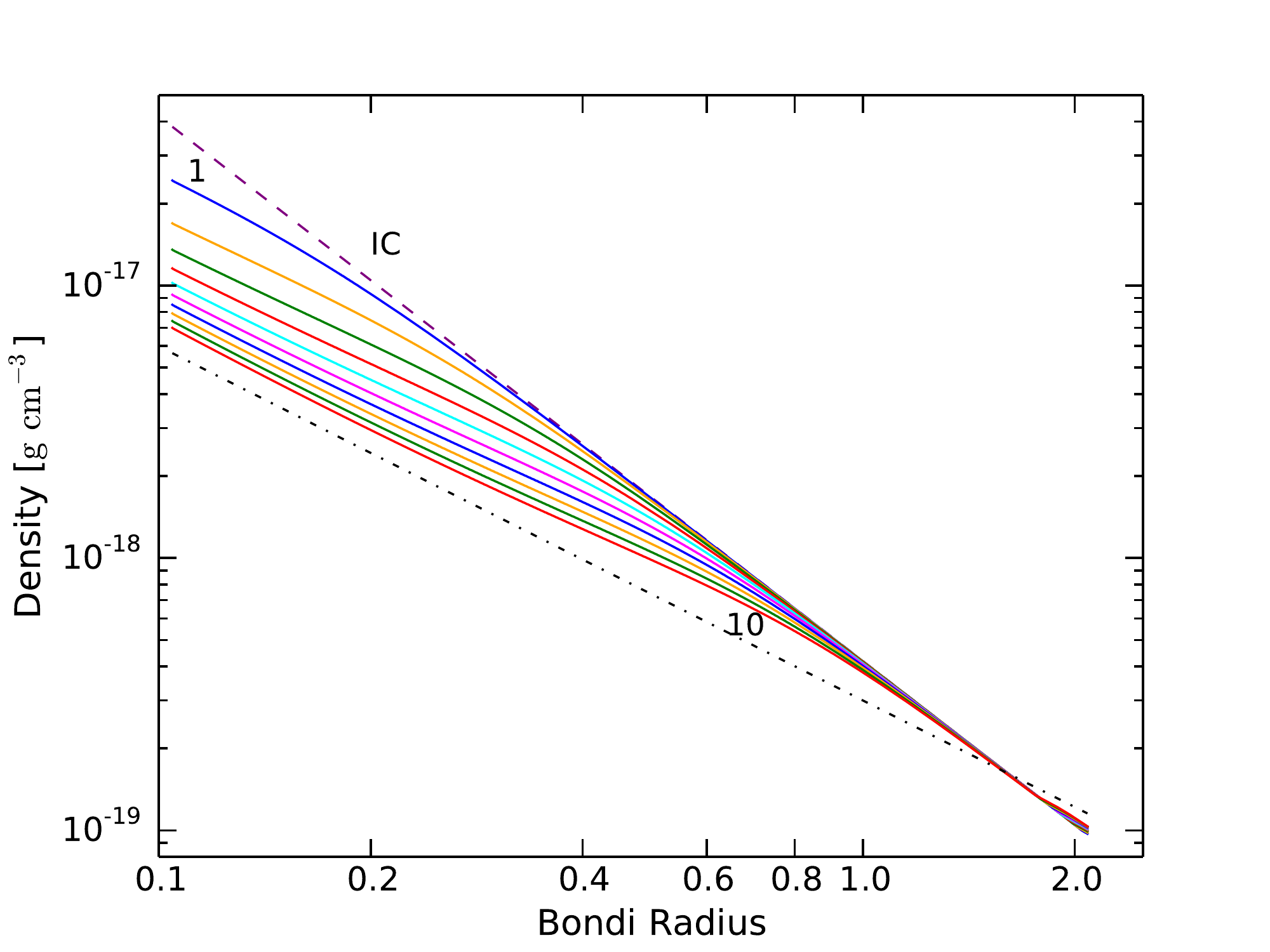}
	\includegraphics[angle=0,width=0.45\textwidth]{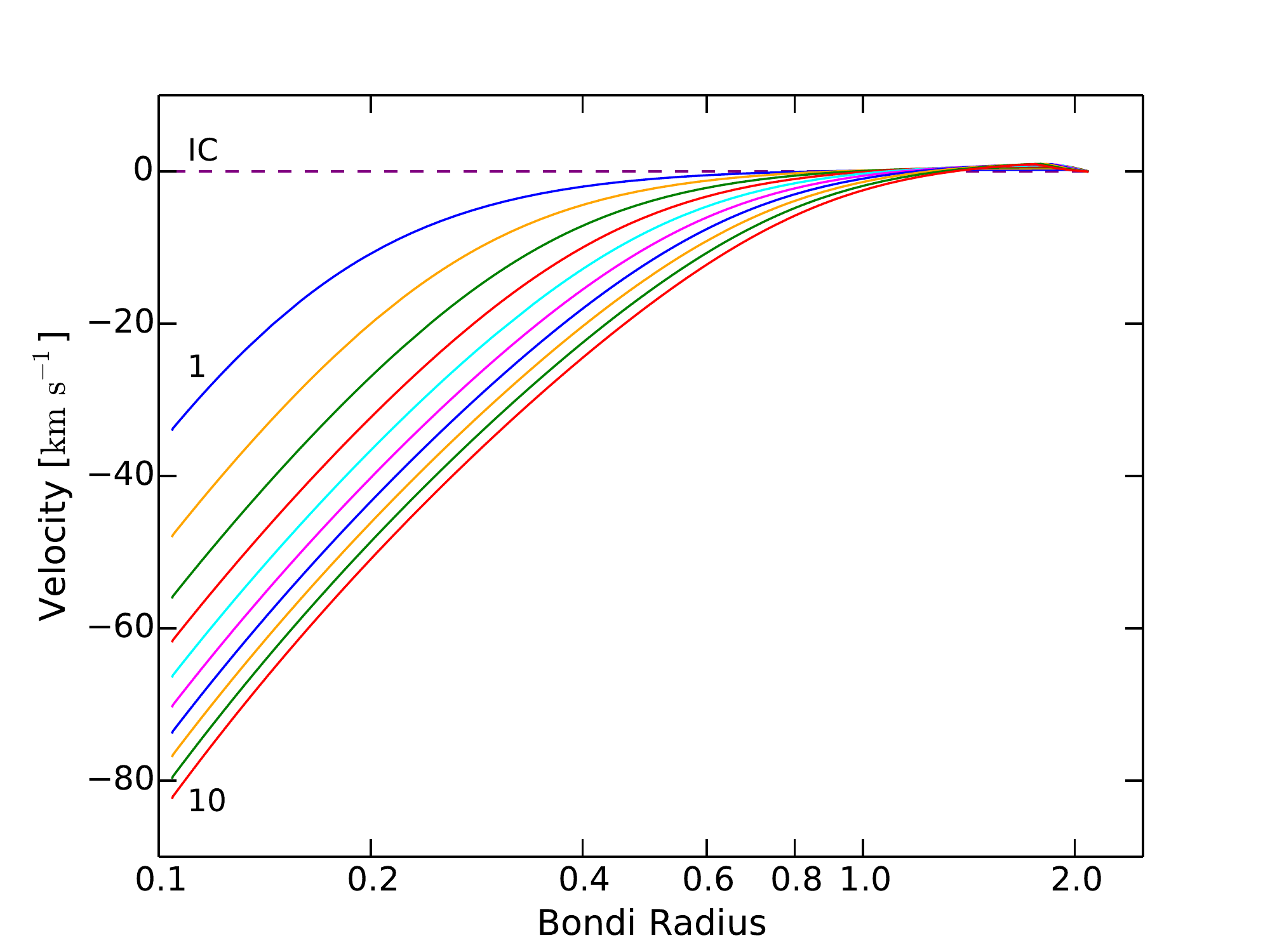}\\
	\vspace{-0.1cm}
	\includegraphics[angle=0,width=0.45\textwidth]{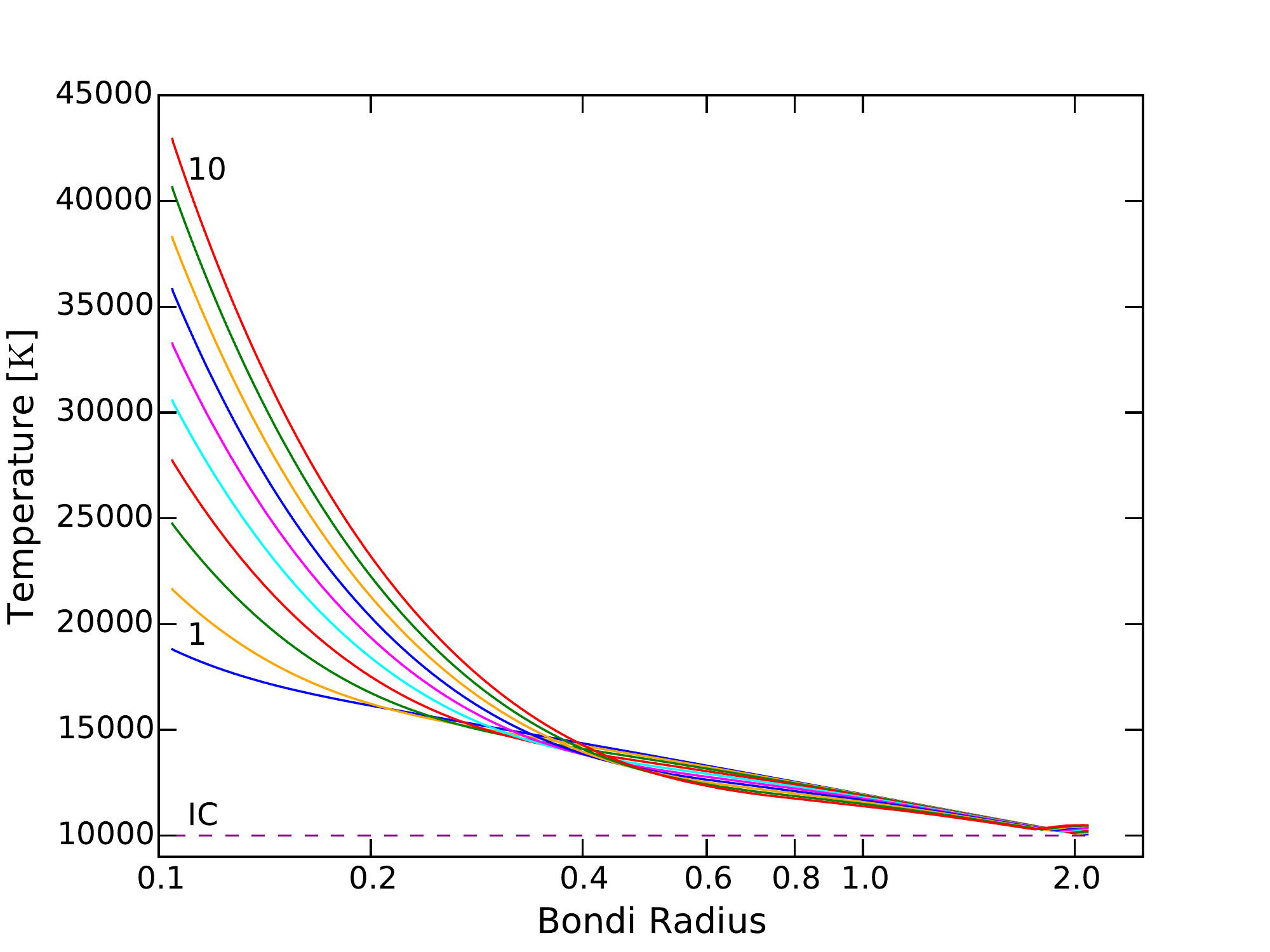}
	\includegraphics[angle=0,width=0.45\textwidth]{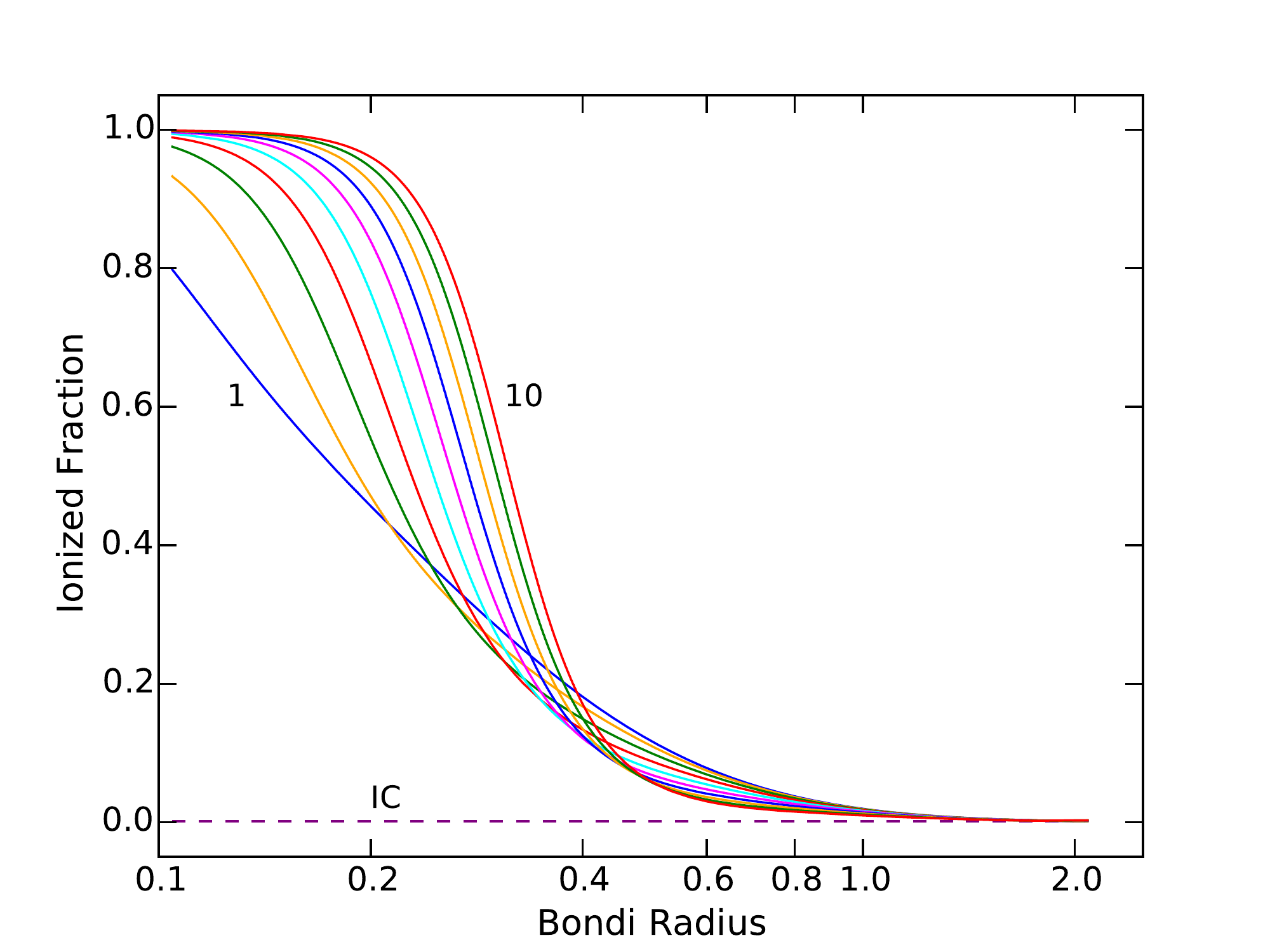}\\
	\vspace{-0.1cm}
	\includegraphics[angle=0,width=0.45\textwidth]{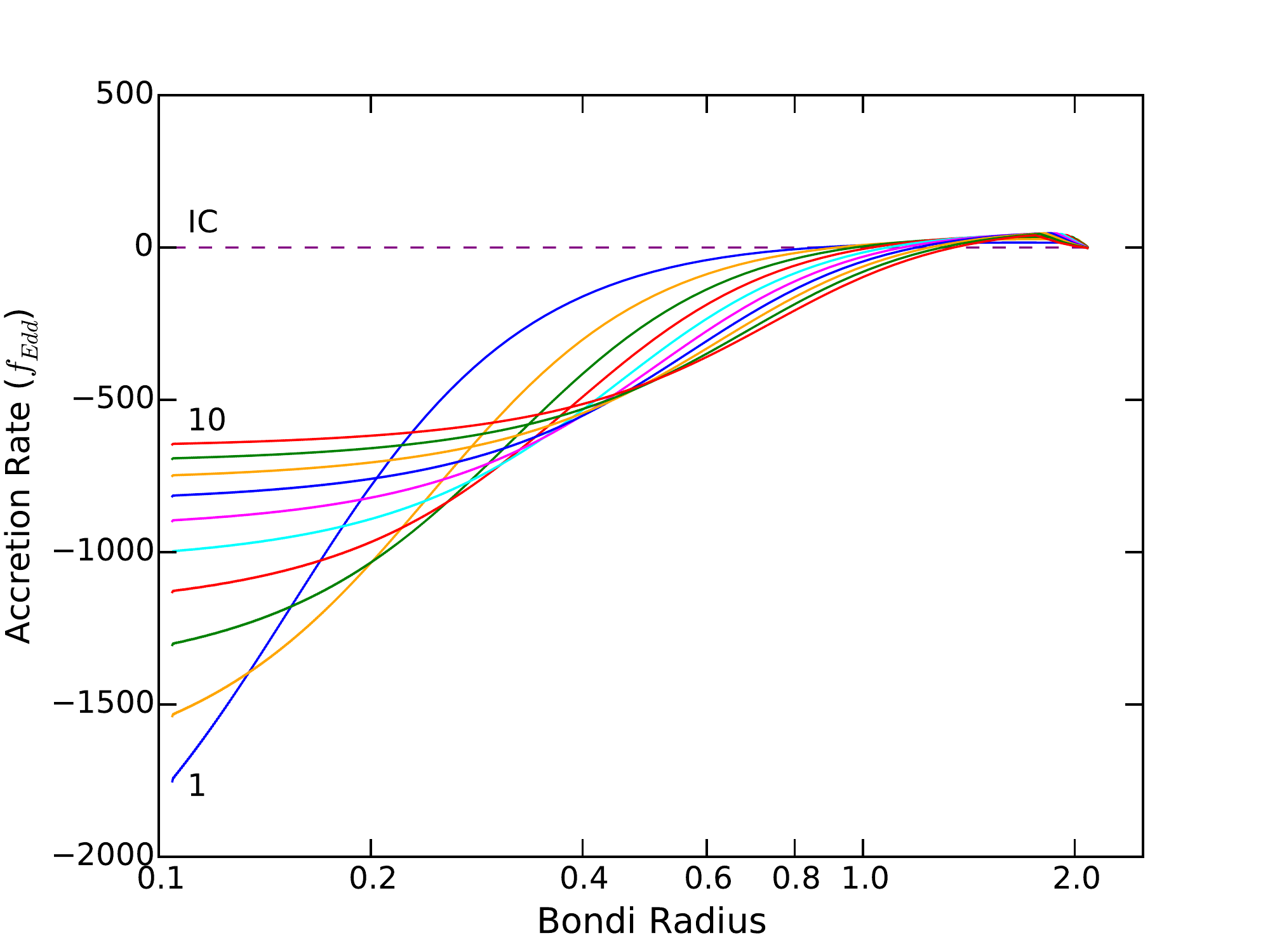}
	\includegraphics[angle=0,width=0.45\textwidth]{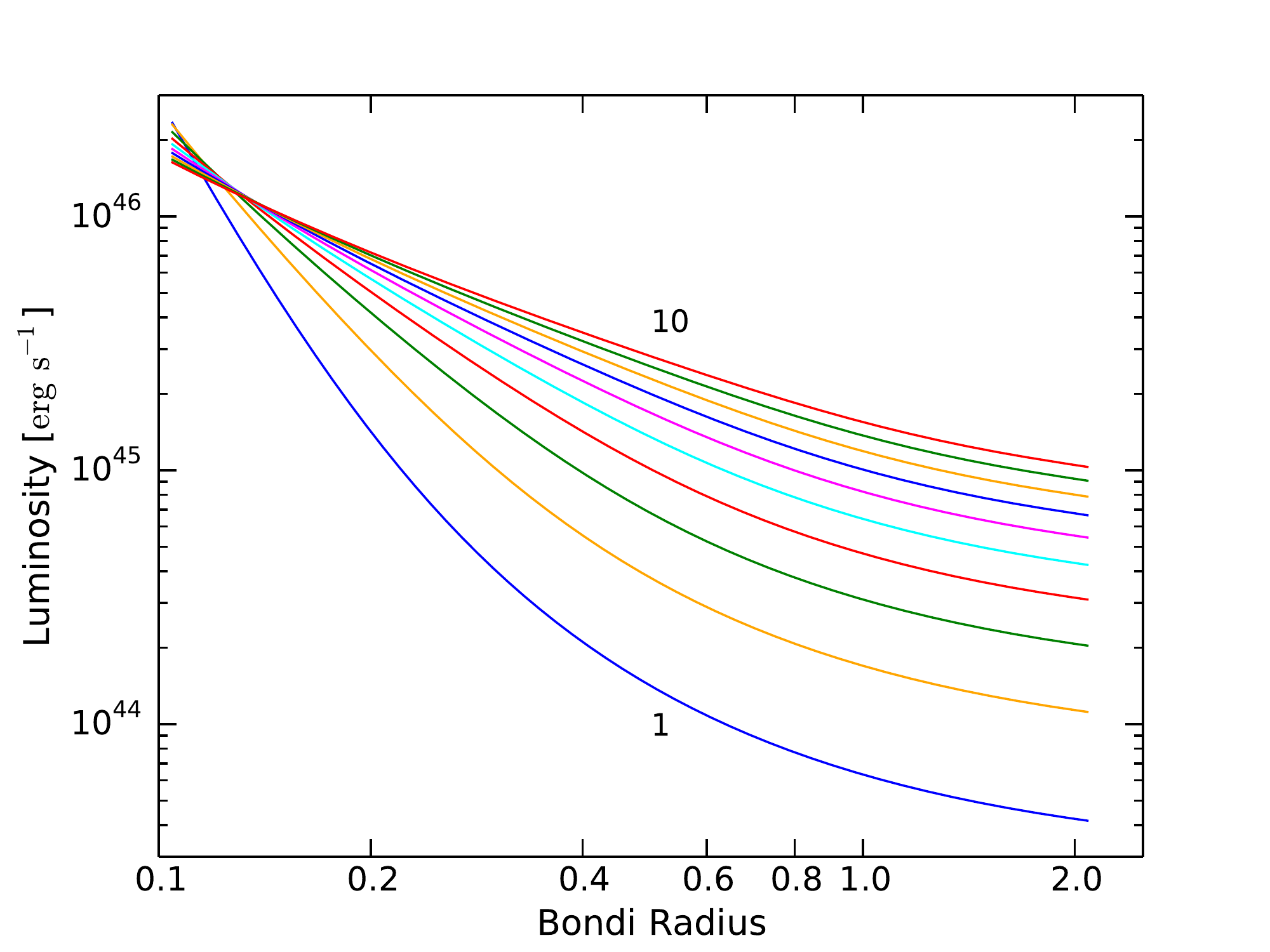}\\
	\caption{Spatial profiles for the T1 simulation: total integration time is $T_{tot} \approx 3.2 \times 10^{4}\, \mathrm{yr}$. The purple line labeled as ``IC" represents the initial conditions: where this line is not present the corresponding initial condition is impossible to show on the plot. The colored lines correspond to different times of the simulation, equally spaced such that $T_{tot} = i \, \Delta t$ with $\Delta t = 3200 \, \mathrm{yr}$ and $i=1, \, ... \, 10$ (for clarity reasons, only the labels $i=1$ and $i=10$ are shown on the plot). All horizontal axes are in units of $R_B$. The density panel reports the classical Bondi solution with a black dot-dashed line. The accretion rate is the mass flux at a given radius and it is plotted with the velocity sign, i.e. a positive (negative) value corresponds to an outflow (inflow).}
	\label{fig:HD_spatial}
\end{figure*}

The evolution occurs over a free-fall time scale, $t_{ff} \sim 3.2 \times 10^4 \, \mathrm{yr}$ during which the system progressively approaches the Bondi solution, reported on the density panel as a black dot-dashed line, in which the gas density scales as:
\begin{equation}
\rho(r) \propto \left( \frac{r}{r_B} \right) ^{-3/2}
\end{equation}
This scaling cannot be sustained for $t \gg t_{ff}$, due to the fact that the gas reservoir is limited (this is not the case in the classical Bondi solution) and ceases to be valid near the Bondi radius, where the gravitational influence of the black hole terminates.

The system is progressively emptied from the inner part of the environment, due to the absence of radiation pressure: the density near the inner boundary drops by a factor $\sim 5$ during the simulation time and the effect is propagated outward, up to the Bondi radius.
The velocity of the gas increases with time as well, stabilizing to a value of order $\sim -100 \, \mathrm{km \, s^{-1}}$ at the inner boundary, which is the result of an equilibrium between the gravitational acceleration and the decreasing thermal pressure of the gas. The temperature of the infalling gas increases, reaching peaks of $4 \times 10^4 \, \mathrm{K}$ in the inner regions. The temperature profile is reflected in the behavior of the ionized fraction (see the Appendix for the relevant equations), which is smaller than $1$ only when the temperature drops below the $\sim 2 \times 10^4 \, \mathrm{K}$ level.
The small jumps visible at the outer boundary of the temperature spatial profile are due to the different rates of change of density and pressure (remind that $T \propto p/\rho$, as in Eq. \ref{EOS}). The gas, outside the sphere of gravitational influence of the black hole, is swept away by the thermal pressure ($dp/dr < 0$ in the entire spatial domain, so that the related force is always in the outward direction) causing an abrupt change in the spatial scaling of pressure and density. This effect is reflected most dramatically on the ionized fraction, since this quantity is very sensitive to modifications of the temperature around the level $\sim 10^4 \, \mathrm{K}$. 
These effects are unimportant for the overall evolution of the system and are visible also in the T2 simulation.

The flow accretes at strongly super-Eddington rates at all times, reaching peaks of $f_{Edd} \equiv \dot{M}_{\bullet}/\dot{M}_{Edd} \sim 2000$: the Eddington limit does not apply in this case due to the absence of radiation pressure. The accretion rate progressively decreases due to the density drop: this causes the slow reduction of the luminosity at the innermost cell, clearly visible in the bottom-right panel of Fig. \ref{fig:HD_spatial}. 

The emitted luminosity is obscured by high column densities at the beginning of the simulation, while the drop of the density decreases the optical depth with time, down to a value $\sim 6$.
The spatial profile of the emitted luminosity is described by Eq. \ref{N1} (and, more specifically, by Eq. \ref{N3} in the Appendix).
The IMBH mass at the end of the T1 simulation is $M_{\bullet} \approx 2.0\times10^5 \, \mathrm{M}_{\odot}$.

\subsubsection{T2: Adding radiation pressure}
In the T2 simulation an outward radiative force, corresponding to a fixed (i.e. not tied to $\dot{M}_{\bullet}$) value of the luminosity $\hat{L} = 2 \times 10^{43} \, \mathrm{erg \, s}^{-1}$, is added to the gravitational force and to the pressure gradient, while the energy equation is still adiabatic. The radiative force is active only when the black hole is accreting, i.e. when $v(r_{min})<0$. The aim of the T2 simulation is to show in a simplified way the effect of the radiation pressure on the gas.

Fig. \ref{fig:luminosity_fix} is a comparison between the fixed value of the luminosity employed in T2 and the luminosity resulting from the T1 simulation, computed through the accretion rate $\dot{M}_{\bullet}$ at $r=r_{min}$ with Eq. \ref{L_BH_definition}, but not included in T1. The latter is $\sim 3$ orders of magnitude larger, due to the absence of radiation pressure quenching the accretion flow. The Eddington luminosity is also shown for comparison, its progressive increase being due to the change in $M_{\bullet}$. The value of $\hat{L}$ is set in order to be at any time larger than the corresponding $L_{Edd}$.

The radiation pressure can modify the accretion flow in two ways.
If $f_{Edd} \equiv \dot{M}_{\bullet}/\dot{M}_{Edd}<1$, the effect is a decrease of the accretion rate $\dot{M}_{\bullet}$. Naming $\dot{M}_{T1}$ the accretion rate without any radiative force (i.e. the one in the T1 simulation) and $\dot{M}_R$ the accretion rate with the addition of a sub-Eddington radiation pressure, it is easy to show that the following relation holds:
\begin{equation}
\dot{M}_R = \dot{M}_{T1} \sqrt{1-f_{Edd}}
\end{equation}
If, instead, $f_{Edd} \geq 1$, the accretion is intermittent (i.e. ${\cal D}<1$): the infalling gas is swept away by the radiation pressure and some time is needed to re-establish the accretion. Under the simplifying assumption that the initial infalling velocity of the gas is slow, it is possible to show that, if $f_{Edd} \geq 1$, an estimate of the value of the duty-cycle is given by:
\begin{equation}
{\cal D} = \left( 2f_{Edd}-1\right)^{-1}
\label{DC_estimate}
\end{equation}
Under these assumptions, ${\cal D} \equiv 1$ for $f_{Edd}=1$, while for $f_{Edd}>1$ it steadily decreases.

The T2 simulation is an example of the latter case, with $f_{Edd} \sim 1.5$. From the previous analysis we expect two major differences with respect to the T1 simulation, namely: (i) the IMBH accretes $\sim 50\%$ less mass (if the total integration times are equal) because ${\cal D}$ is smaller by a factor $\sim 2$,  and (ii) the IMBH produces some feedback effect on the surrounding gas.
This is exactly what we observe in the T2 simulation. The final mass is $M_{\bullet} = 1.4\times10^5 \, \mathrm{M}_{\odot}$, i.e. the black hole accreted $\sim 60\%$ less mass than in the T1 simulation, in agreement with our rough estimate in Eq. \ref{DC_estimate}.
In addition, the spatial profiles for this simulation, shown in Fig. \ref{fig:RAD_FORCE_spatial}, manifestly provide some evidence of the effect that the radiation pressure exerts on the gas.
A density wave, produced by the radiative feedback, propagates in the outward direction with velocities up to $\sim 20 \, \mathrm{km\, s}^{-1}$. It is interesting to note that this wave is mildly supersonic, since the value of the sound speed in a gas at $T \sim 2.6 \times 10^4 \, \mathrm{K}$ is $\sim 19 \, \mathrm{km \, s^{-1}}$. In addition, the positive values of the accretion flow measured in a large part of the spatial domain indicate the occurrence of a gas outflow from the external boundary. The high-density wave, with temperatures as high as $\sim 2.8 \times 10^4 \, \mathrm{K}$, is followed by a rarefaction zone, where the temperature drops to $\sim 7000 \, \mathrm{K}$ (the temperature profile follows the pressure, i.e. the cooling is adiabatic), decreasing the ionized fraction to very small ($\sim 10^{-3}$) values as well.
The accretion flow promptly (after $\sim 6000 \, \mathrm{yr}$) stabilizes to a value $f_{Edd} \sim -500$ at the innermost cell: this value is set by the fixed radiation pressure of the T2 simulation, which indirectly sets also the velocity at which the accretion flow is re-established at the end of each idle phase (the larger is the radiation pressure, the longer is the time needed for accretion to re-start, then the larger is the resulting mass inflow). 

This general framework is explained by the following scenario: when the black hole accretes, the fixed super-critical emitted luminosity sweeps away the surrounding gas, affecting a spherical region of radius $r_{\tau}$, defined such that $\tau(r_{\tau}) = 1$. The gas located at $r \gg r_{\tau}$ is also accelerated upward, not by the radiation pressure in this case but by the thermal pressure, and acquires a positive velocity. When the irradiation is temporarily shut down, the gas located at $r \ll r_{\tau}$ is affected by the strong gravitational field of the black hole and falls back in due course. 

\begin{figure}
\vspace{-1\baselineskip}
\hspace{-0.5cm}
\begin{center}
\includegraphics[angle=0,width=0.45\textwidth]{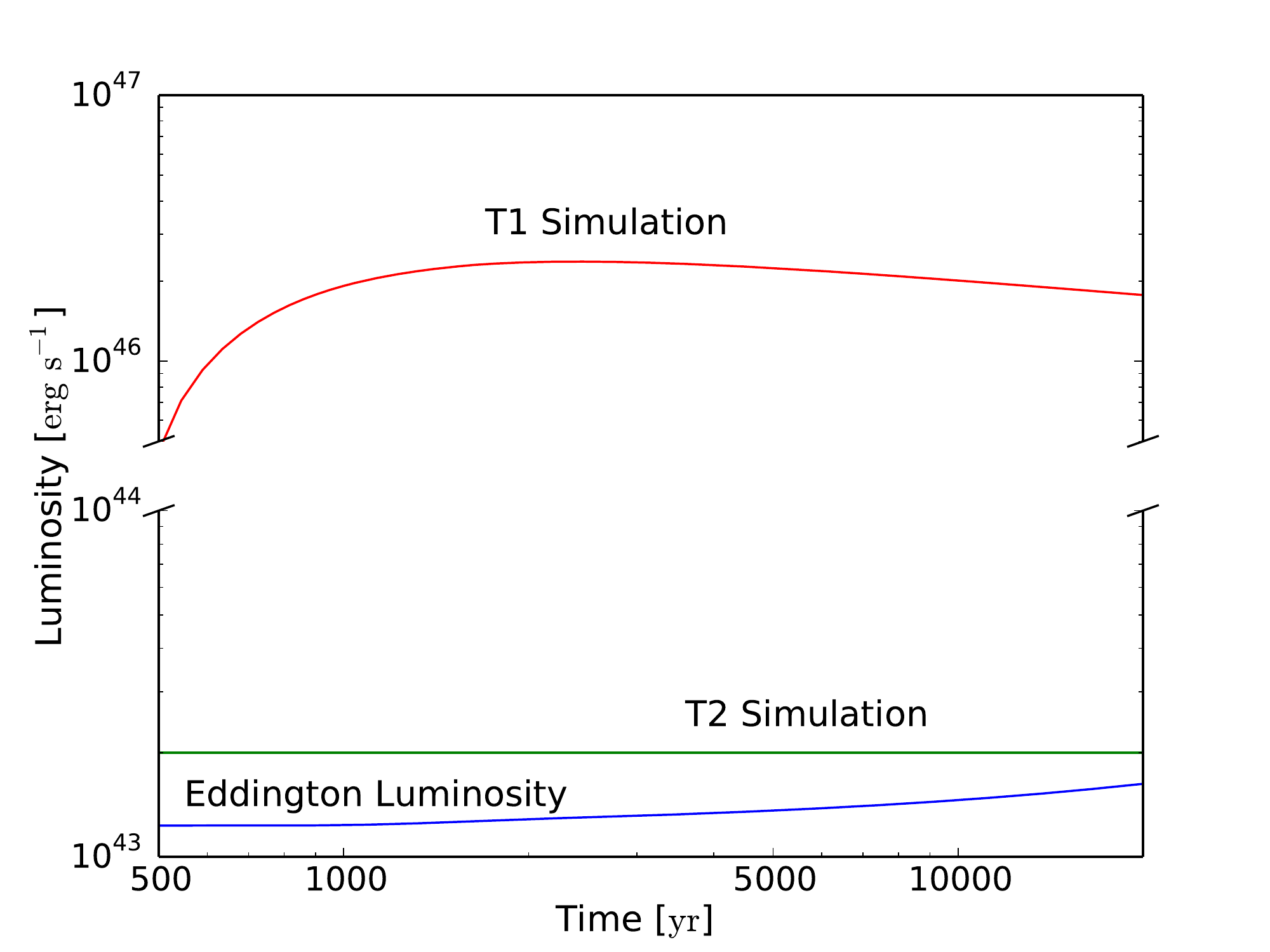}
\caption{Comparison between the emitted luminosity for different simulations (note that the vertical axis is broken). The red line is the luminosity of the T1 simulation (computed from $\dot{M}_{\bullet}$ at $r=r_{min}$ with Eq. \ref{L_BH_definition}, but not included in the physics). In green, the fixed luminosity value used for the T2 simulation $\hat{L}=2\times10^{43} \, \mathrm{erg \, s}^{-1}$. In blue, for comparison, the Eddington luminosity for the T2 simulation, whose value increases along with $M_{\bullet}$. The value of $\hat{L}$ is always above the Eddington limit.}
\label{fig:luminosity_fix}
\end{center}
\end{figure}

\begin{figure*}
	\includegraphics[angle=0,width=0.45\textwidth]{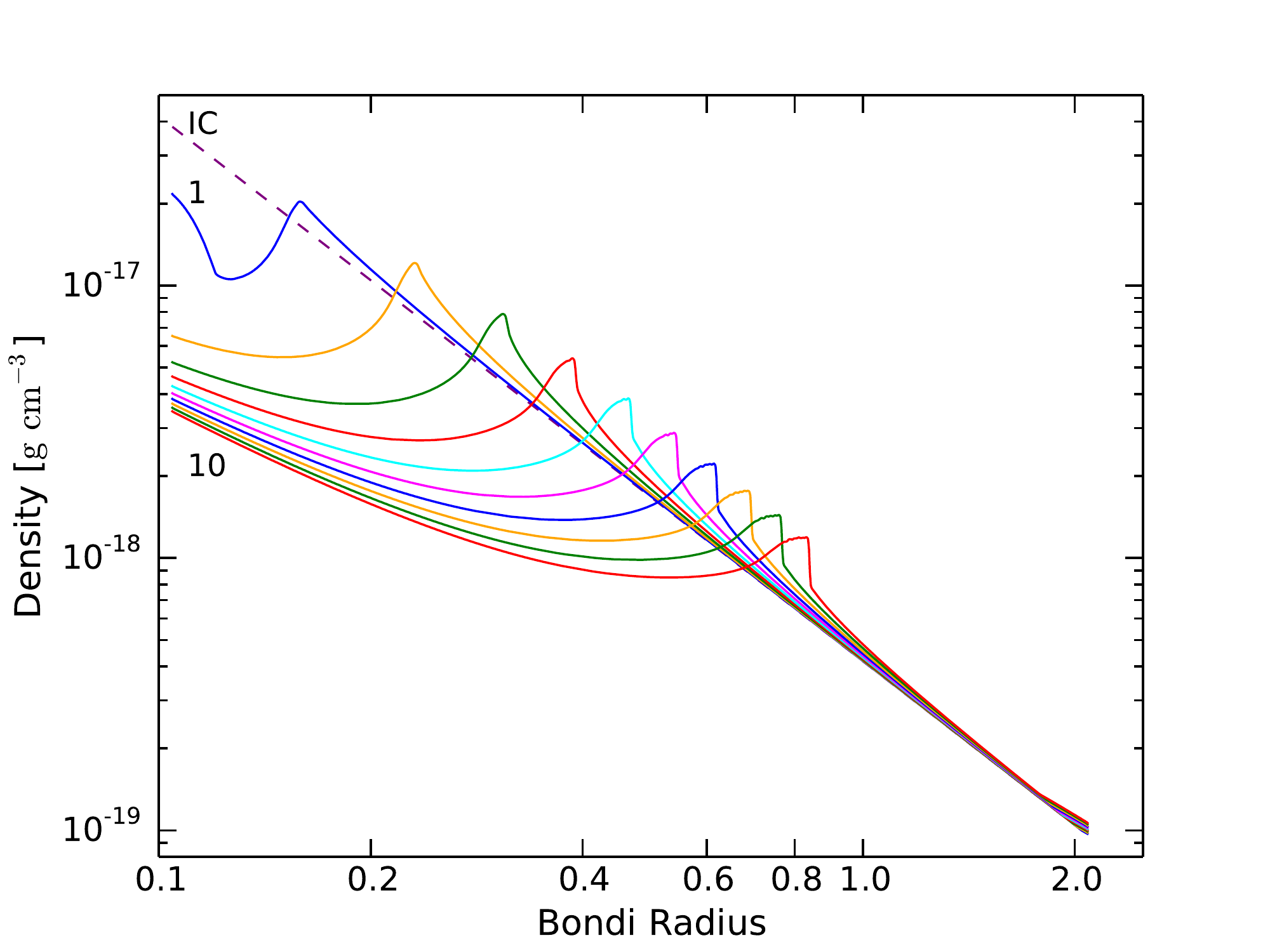}
	\includegraphics[angle=0,width=0.45\textwidth]{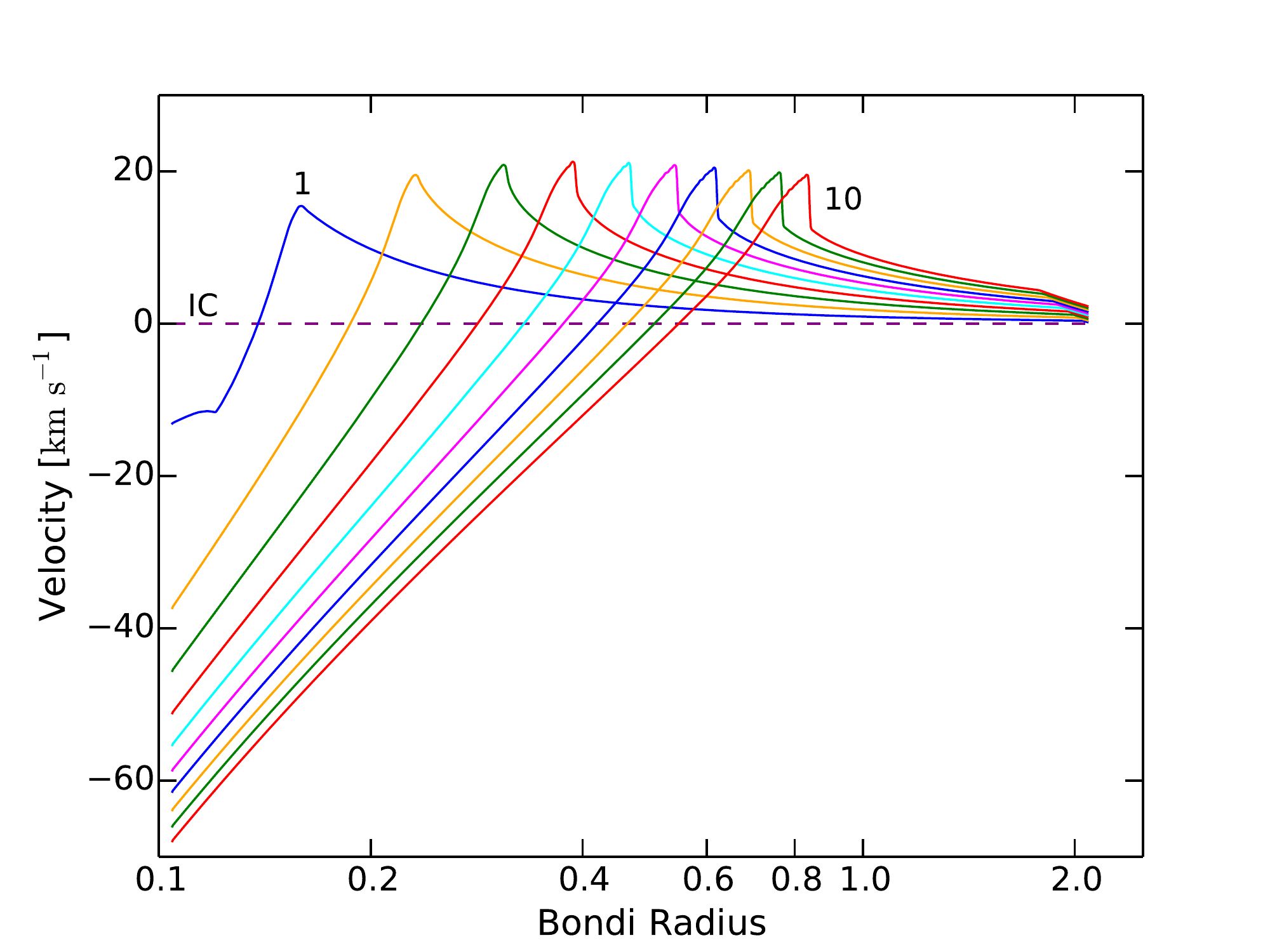}\\
	\vspace{-0.1cm}
	\includegraphics[angle=0,width=0.45\textwidth]{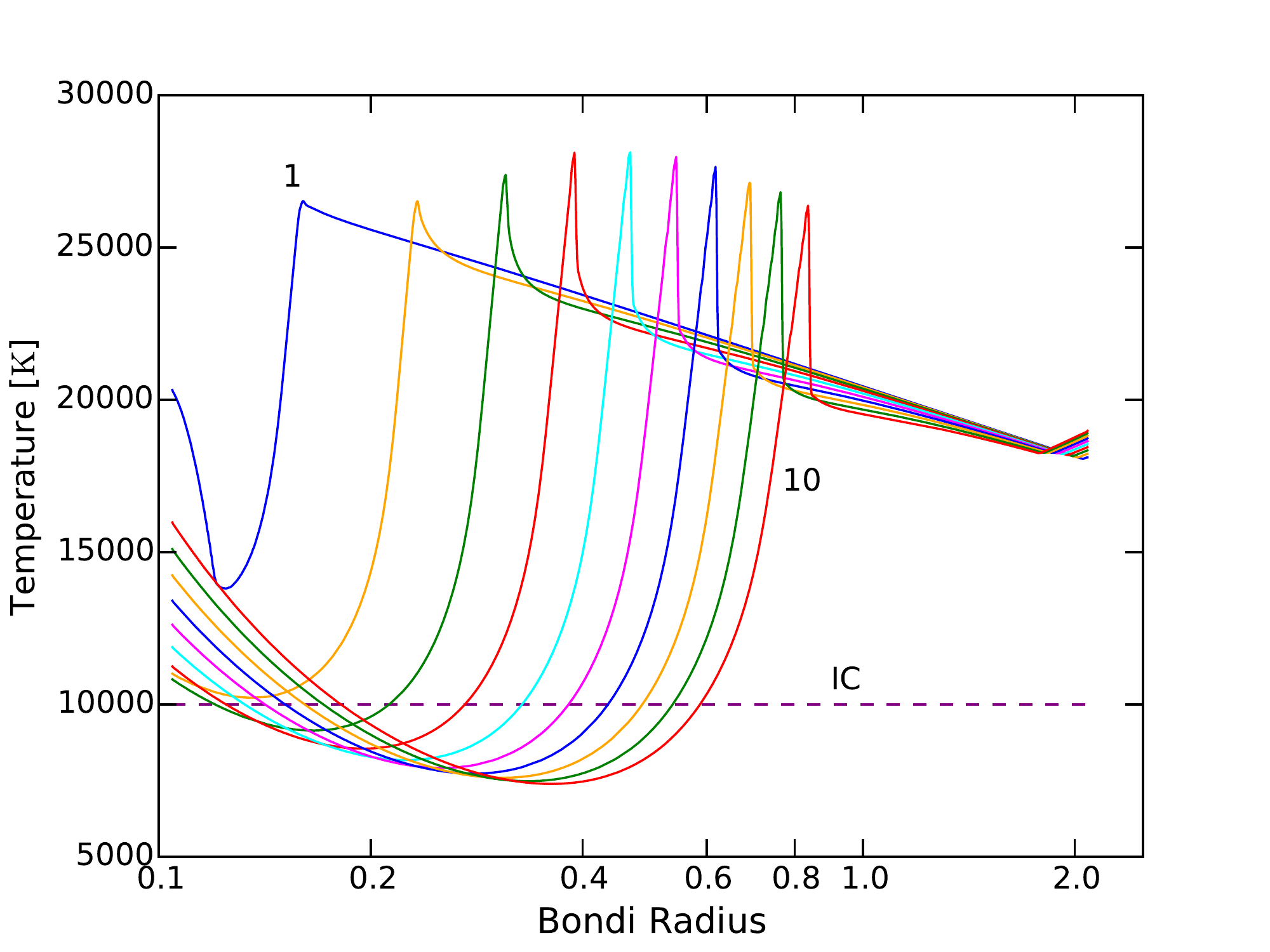}
	\includegraphics[angle=0,width=0.45\textwidth]{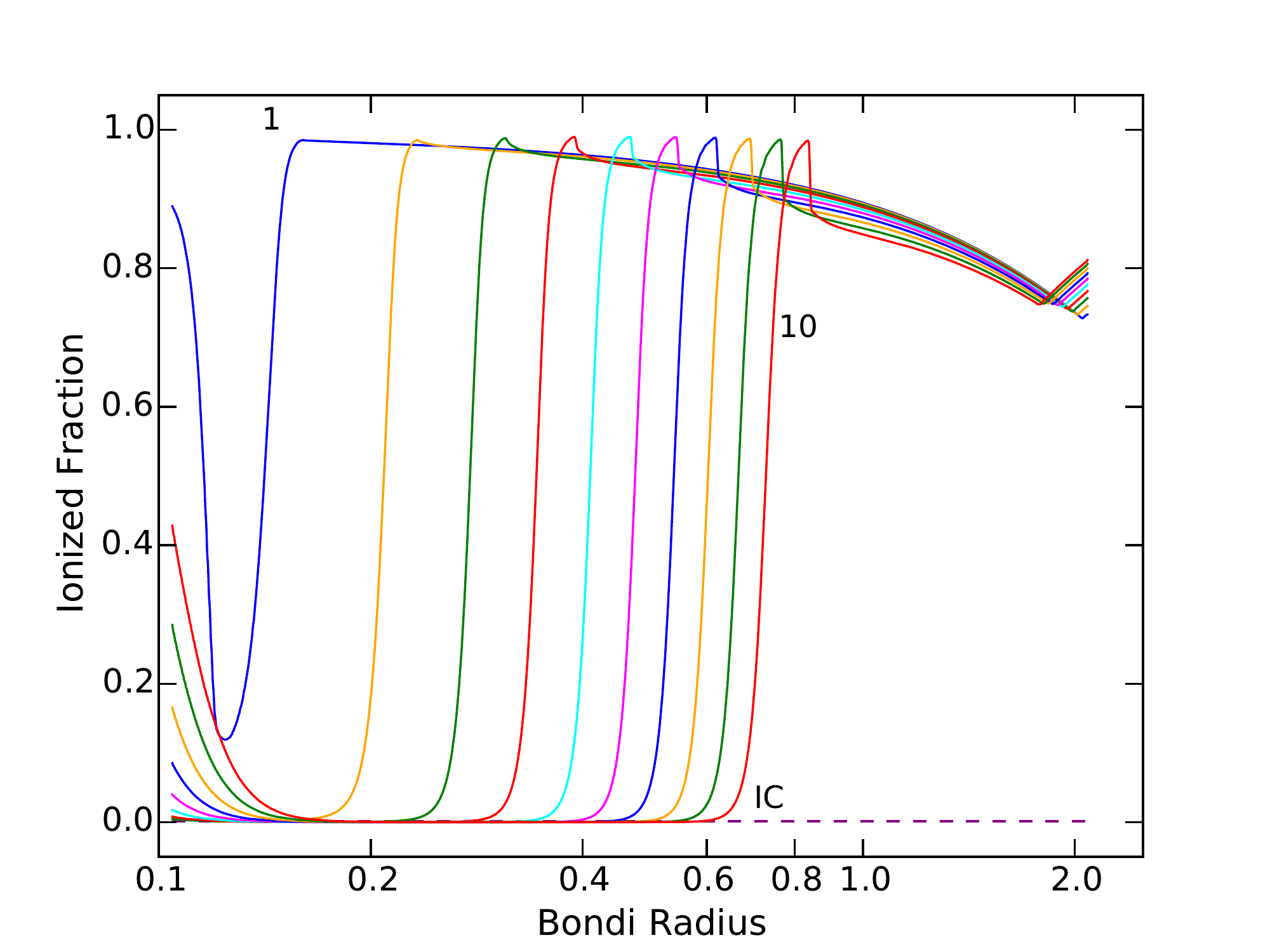}\\
	\vspace{-0.1cm}
	\includegraphics[angle=0,width=0.45\textwidth]{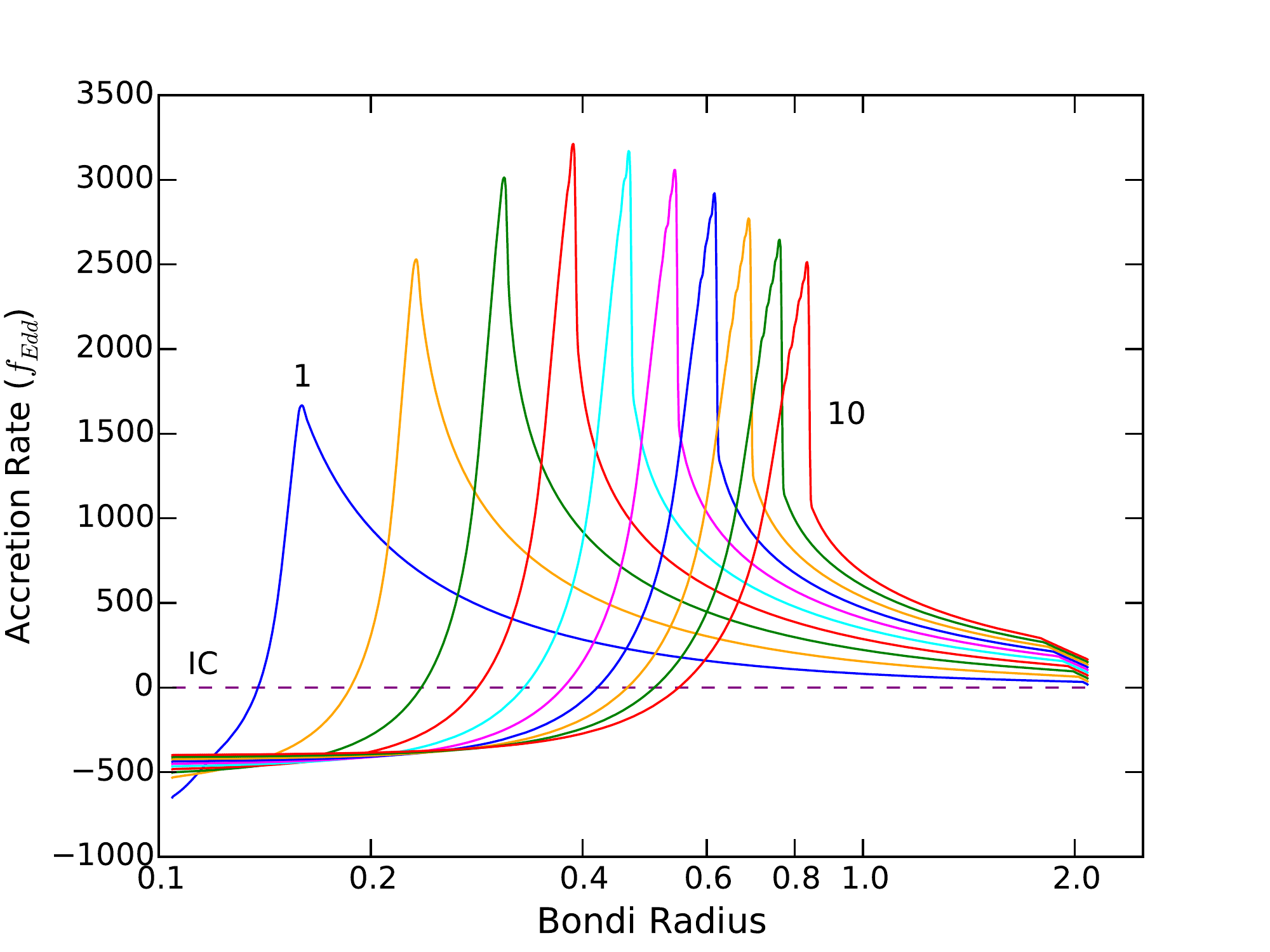}
	\includegraphics[angle=0,width=0.45\textwidth]{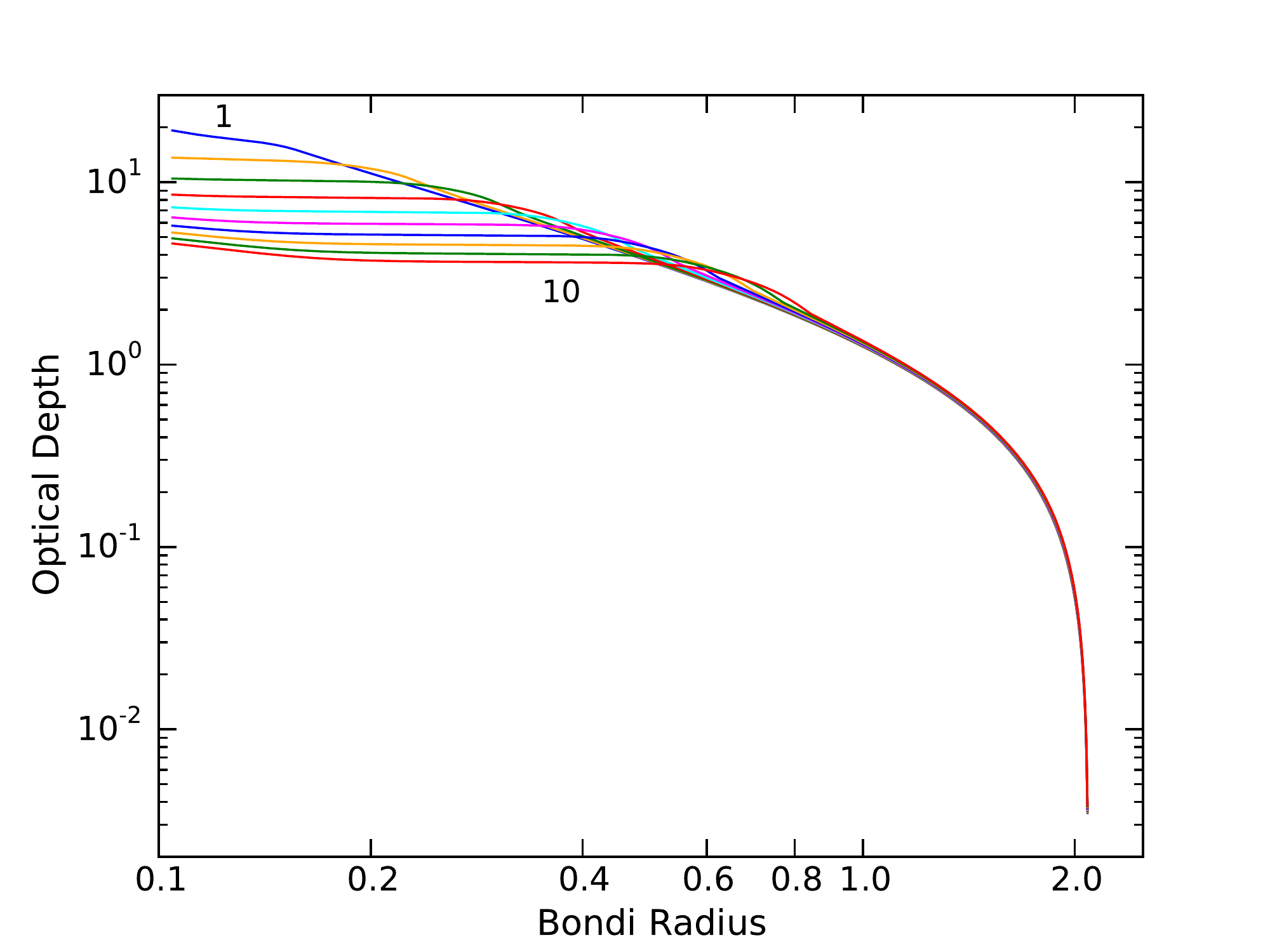}\\
	\caption{Spatial profiles for the T2 simulation: total integration time is $T_{tot} \approx 3.2 \times 10^{4}\, \mathrm{yr}$. A fixed radiative force, determined by the luminosity $\hat{L}=2\times 10^{43} \, \mathrm{erg \, s}^{-1}$ and active only when $v(r_{min})<0$, is added to the gravity and to the pressure gradient, as explained in the text. The colored lines correspond to different times of the simulation, equally spaced such that $T_{tot} = i \, \Delta t$ with $\Delta t = 3200 \, \mathrm{yr}$ and $i=1, \, ... \, 10$ (for clarity reasons, only the labels $i=1$ and $i=10$ are shown on the plot).}
	\label{fig:RAD_FORCE_spatial}
\end{figure*}

\subsection{The Full Simulation}
The aim of the FS simulation is to describe the accretion flow onto a DCBH with initial mass $M_0 = 10^5 \, \mathrm{M_{\odot}}$. The final integration time for this simulation, when all the gas contained in the halo is expelled by radiation pressure, is $\sim 142 \, \mathrm{Myr}$.
The forces acting on the gas are the gravity of the black hole, the pressure gradient and the radiation pressure. 

The differences with respect to the previous T1 and T2 simulations are: (i) the radiation pressure is computed self-consistently, i.e. with Eq. \ref{L_BH_definition} and (ii) the energy equation is non-adiabatic, with the inclusion of bremsstrahlung cooling and atomic cooling (see the Appendix).

\subsubsection{Spatial structure and time evolution}
Broadly speaking, the simulation allows the identification of three distinct evolutionary phases of the system, described in turn below.
\begin{enumerate}
\item{\textbf{Initial Transient Phase}: the gas, initially almost at rest (as detailed in Sec. \ref{sec:methods}), is accelerated downward, progressively increasing $\dot{M}_{\bullet}$ and, as a consequence, the emitted luminosity $L_{\bullet}$, as shown in Fig. \ref{fig:luminosity_trend_start}. This plot shows that the emitted luminosity increases by $\sim 3$ orders of magnitude in only $\sim 200 \, \mathrm{yr}$, a fraction $\sim 10^{-6}$ of the full evolution of the system.
This process is self-regulated, due to the interconnection between gravity, accretion rate and radiation pressure: the gravity tends to increase the accretion rate by accelerating the gas downward, while the emitted luminosity acts against the infall by providing an outward acceleration.
This evolutionary phase lasts until the emitted luminosity becomes comparable to the Eddington limit (see Eq. \ref{eq:L_edd}), approximately $10^{43} \, \mathrm{erg \, s}^{-1}$ at the beginning of the simulation. 
Above this threshold, the radiation pressure is able to sweep the gas away from the inner boundary and the accretion process becomes intermittent (${\cal D}< 1$).

This initial phase lasts $\sim 200-300 \, \mathrm{yr}$: the emitted luminosity and the fractional mass accreted ($\Delta M_t/M_0 = (M_t-M_0)/M_0$) are shown in Fig.\ref{fig:luminosity_trend_start}.
An estimate of the duration $T_t$ of this initial phase is easily computed from the following argument.
We request that during $T_t$ the luminosity $L_{\bullet}=\eta c^2 \dot{M}_{\bullet}$ becomes equal to the Eddington luminosity:
\begin{equation}
L=\eta c^2 \frac{dM}{dt} \equiv L_{Edd} = \frac{4 \pi G m_p c}{\sigma_T} M
\end{equation}
By means of integrating between $t=0$ when $M(t) = M_0$ and $T_t$ when $M(t) = M_t$ and calling $\Delta M_t = M_t-M_0$ we obtain:
\begin{equation}
T_t = \frac{\eta c \sigma_T}{4 \pi G m_p} \ln \left( 1 + \frac{\Delta M_t}{M_0} \right) \approx \frac{\eta c \sigma_T}{4 \pi G m_p} \frac{\Delta M_t}{M_0}
\end{equation}
where the last approximation is valid for $\Delta M_t/M_0 \ll 1$.
This equation, with the value $\Delta M/M_0 \approx 7 \times 10^{-6}$ taken from Fig. \ref{fig:luminosity_trend_start}, gives the expected time scale $T_t \sim 300 \, \mathrm{yr}$.
Defining $t_{Edd} \equiv (c \sigma_T)/(4 \pi G m_p)$ the Eddington time scale, the previous formula becomes:
\begin{equation}
T_t \approx \eta t_{Edd} \frac{\Delta M}{M_0}
\end{equation}
Interestingly, if we suppose that $M(t) \propto M_0$ (see e.g. \citealt{Volonteri_2014, Madau_2014}), as in:
\begin{equation}
M(t) = M_0 \, \exp{ \left[ \left( \frac{1-\eta}{\eta} \right) \frac{t}{t_{Edd}} \right] } 
\end{equation} 
the time scale $T_t$ is strictly independent on the initial black hole mass $M_0$.

It is relevant to investigate in detail the mechanism which leads the system from a stable accretion at $L \sim L_{Edd}$ to the unstable and intermittent phase shown in Fig. \ref{fig:luminosity_trend_start} for $t \gtrsim 270 \, \mathrm{yr}$.
When the black hole starts to accrete at $\dot{M}_{\bullet} \sim \dot{M}_{Edd}$, the gas near the inner boundary is swept away, so that the accretion is temporarily interrupted and the radiation pressure is turned off. The emitted luminosity does not affect the outer parts of the domain, for $r \gg r_{\tau}$, see the similar discussion for the T2 simulation. The gas located in this section continues its infall, counteracted only by the thermal pressure exerted from the internal layers, and eventually feeds the black hole with increasingly larger mass inflows. The difference with respect to the T2 simulation is due to the direct dependence of the emitted luminosity from the mass inflow in the FS case. More specifically, this mechanism is schematically explained in Fig. \ref{fig:mechanism}.
In the first panel, the black hole is accreting mass from the innermost shell, which is collapsing just as the outer one. When the black hole irradiates with $L \gtrsim L_{Edd}$, the radiation pressure acts only on the innermost shell (supposing for simplicity that the outer shell is located at $r \gg r_{\tau}$) which is swept outward, while the outer shell continues the infall. During this period, the black hole is not irradiating. The innermost shell eventually terminates its outward movement, due to the gravitational pull of the black hole. The innermost shell merges with the outer one, so that its overall density is increased. Eventually, the merged shells approach the accretion boundary and re-start the cycle with a larger accretion flow, i.e. with the emission of a higher luminosity.

\begin{figure}
\vspace{-1\baselineskip}
\hspace{-0.5cm}
\begin{center}
\includegraphics[angle=0,width=0.5\textwidth]{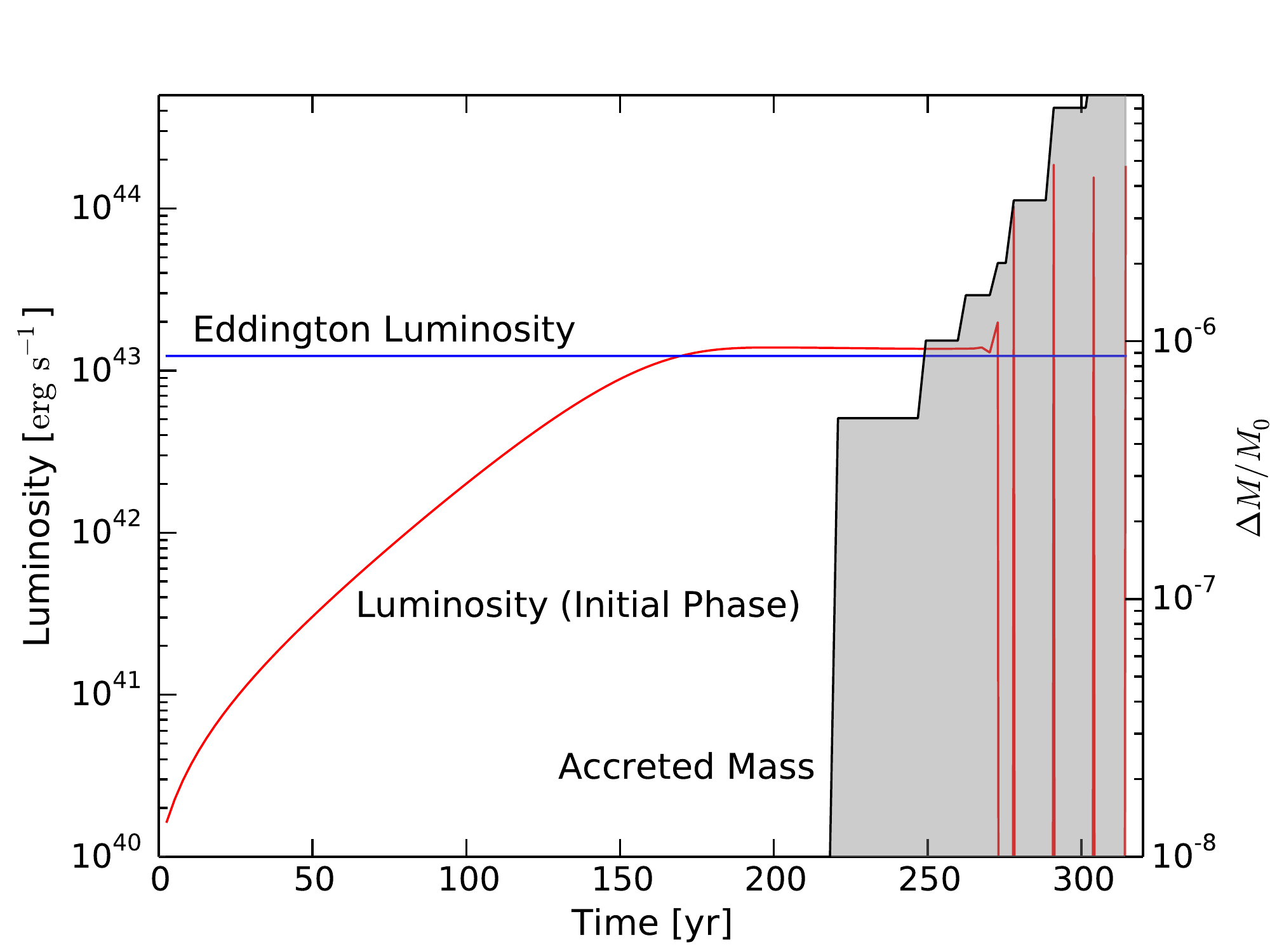}
\caption{Luminosity emitted and fractional mass accreted ($\Delta M_t /M_0 =(M_t-M_0)/M_0$) during the initial phase, lasting $\sim 200-300 \, \mathrm{yr}$. The corresponding Eddington luminosity is also shown, for comparison.}
\label{fig:luminosity_trend_start}
\end{center}
\end{figure}

\begin{figure}
\vspace{-0.5\baselineskip}
\hspace{-0.5cm}
\begin{center}
\includegraphics[angle=0,width=0.45\textwidth]{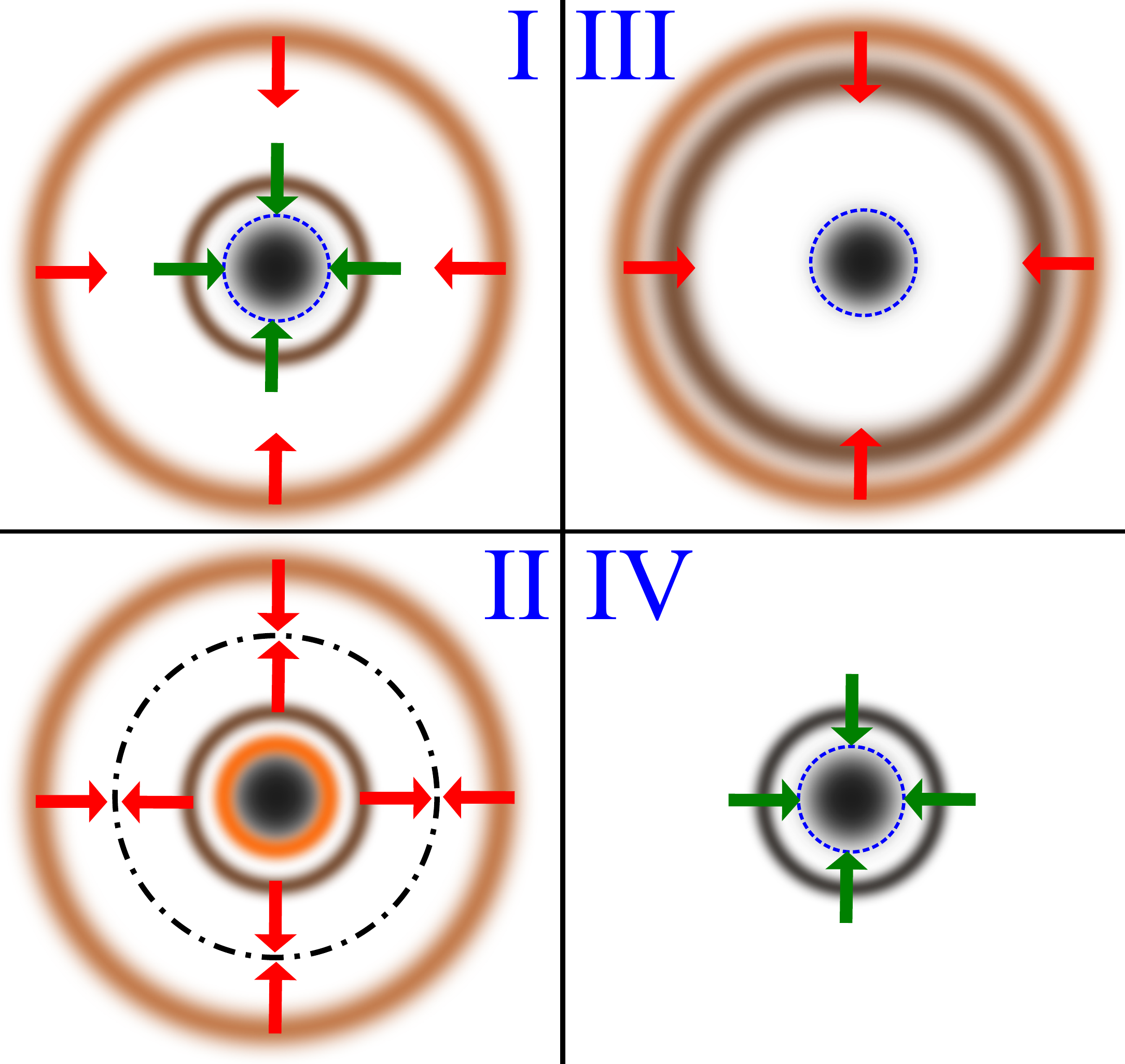}
\caption{Schematic description of the mechanism which progressively increases the emitted luminosity. Here we consider only two mass shells, but the system must be thought to be composed of an infinite number of them. The black hole is at the center of each of the four panels, in black. Brown circles are mass shells: the darker the shell, the higher the density. The smallest circle is the accretion boundary: it becomes orange when the black hole irradiates. Green arrows indicate accretion through the inner boundary, red arrows indicate a movement of the mass shell. The black dot-dashed line indicates the radius $r_{\tau}$. See the text for a detailed description of the panels.}
\label{fig:mechanism}
\end{center}
\end{figure}
}

\item{\textbf{Main Accretion Phase}: this phase lasts $\sim 142 \, \mathrm{Myr}$ and is characterized by a duty-cycle ${\cal D} \sim 0.48$ and an average accretion rate $f_{Edd} \simeq 1.35 $: the accretion is super-critical on average. This value is computed as a global average of $f_{Edd}$ over the entire integration time, including the idle phases (when the accretion does not take place) and it is in substantial agreement with the approximated (i.e. valid for small inflow velocities) relation given in Eq. \ref{DC_estimate}.

\begin{figure*}
	\includegraphics[angle=0,width=0.45\textwidth]{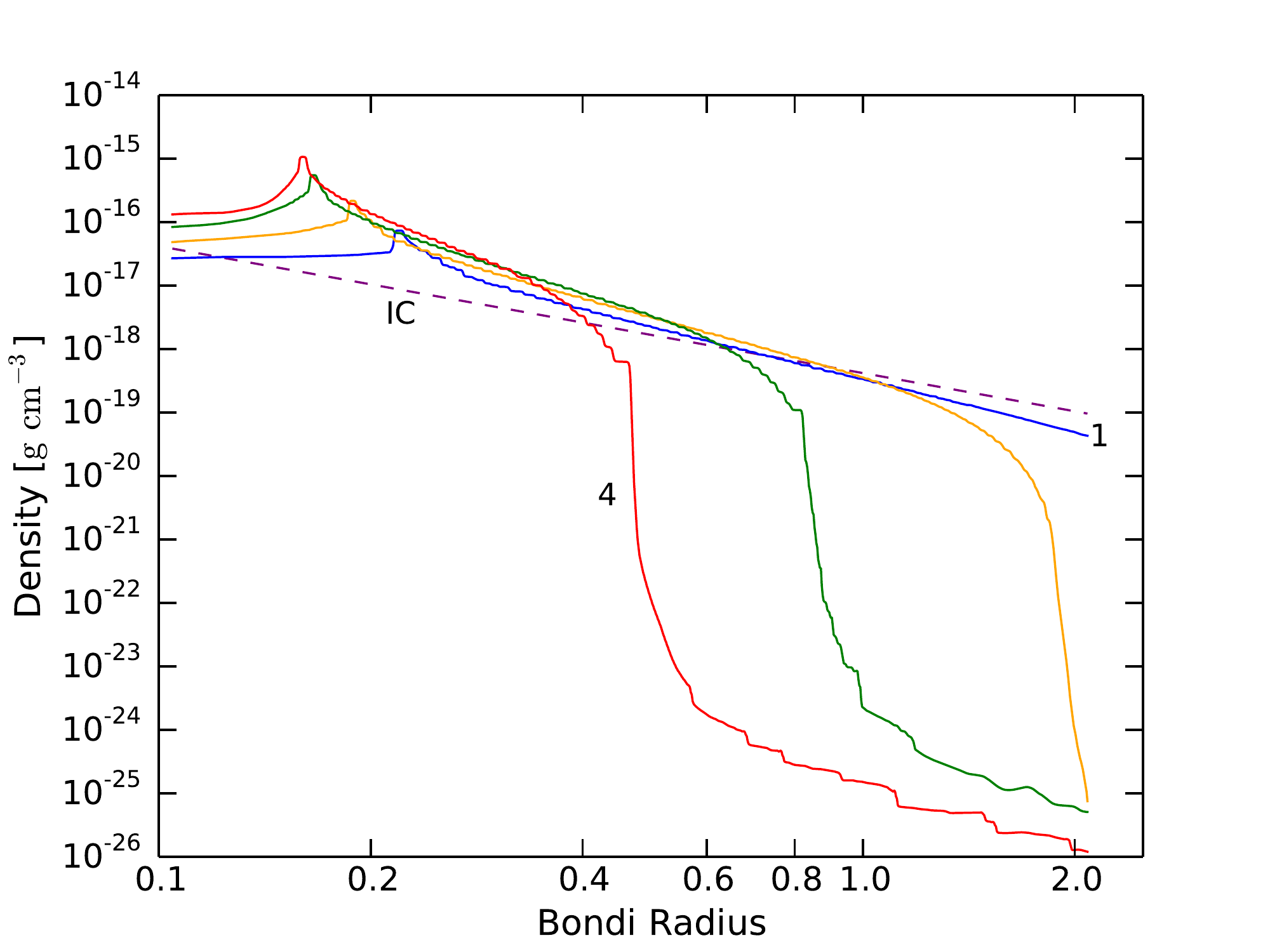}
	\includegraphics[angle=0,width=0.45\textwidth]{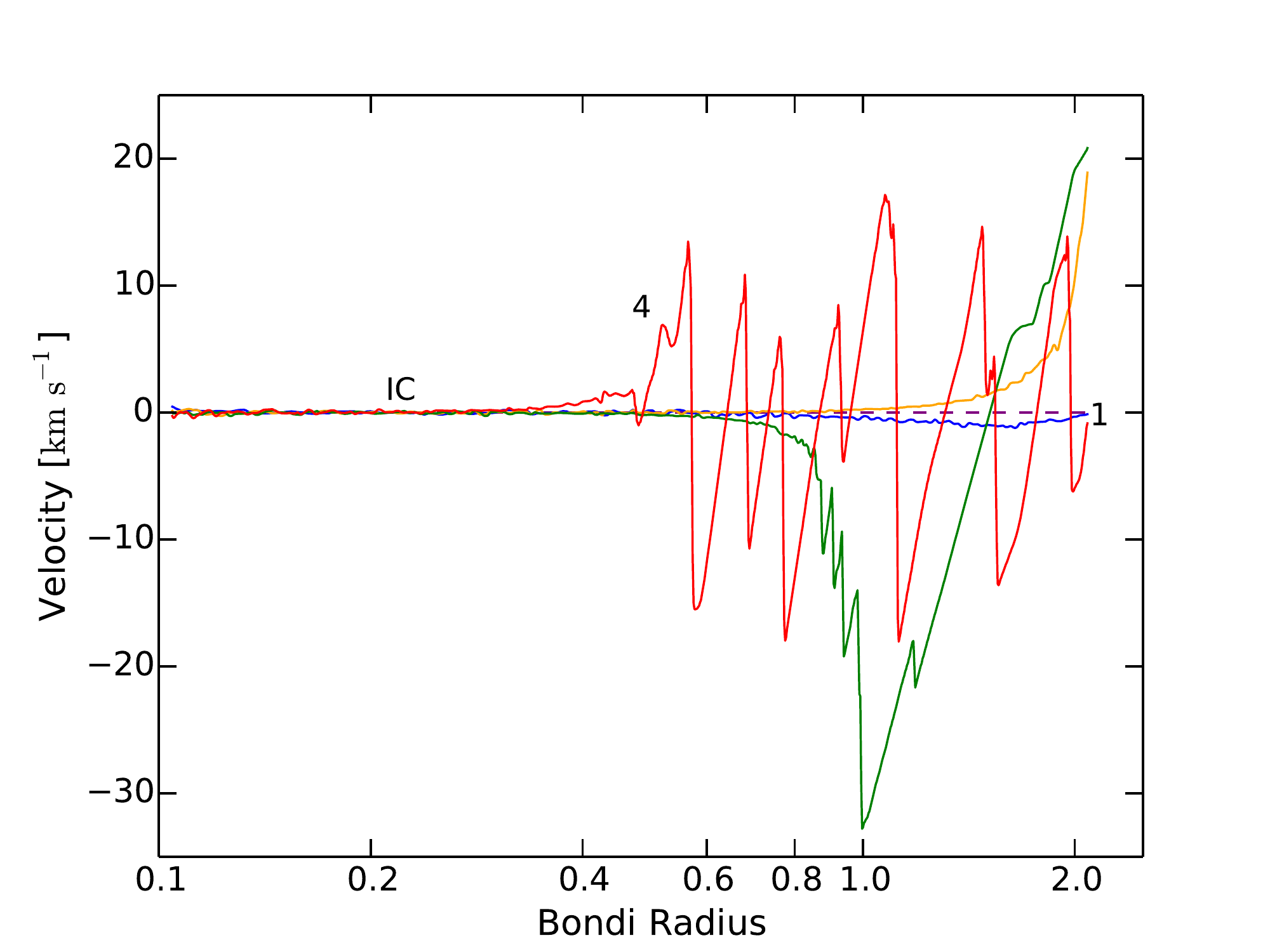}\\
	\vspace{-0.1cm}
	\includegraphics[angle=0,width=0.45\textwidth]{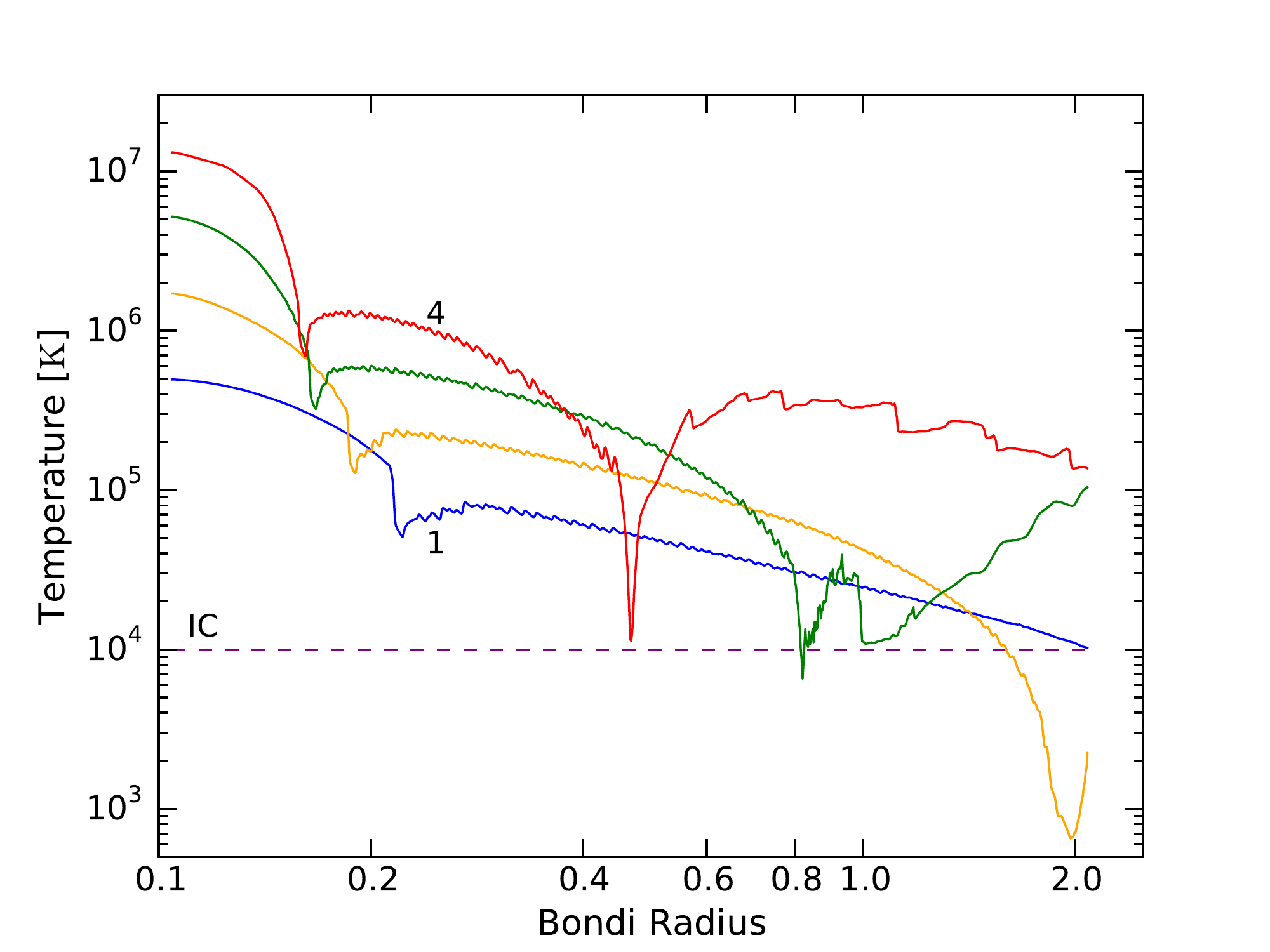}
	\includegraphics[angle=0,width=0.45\textwidth]{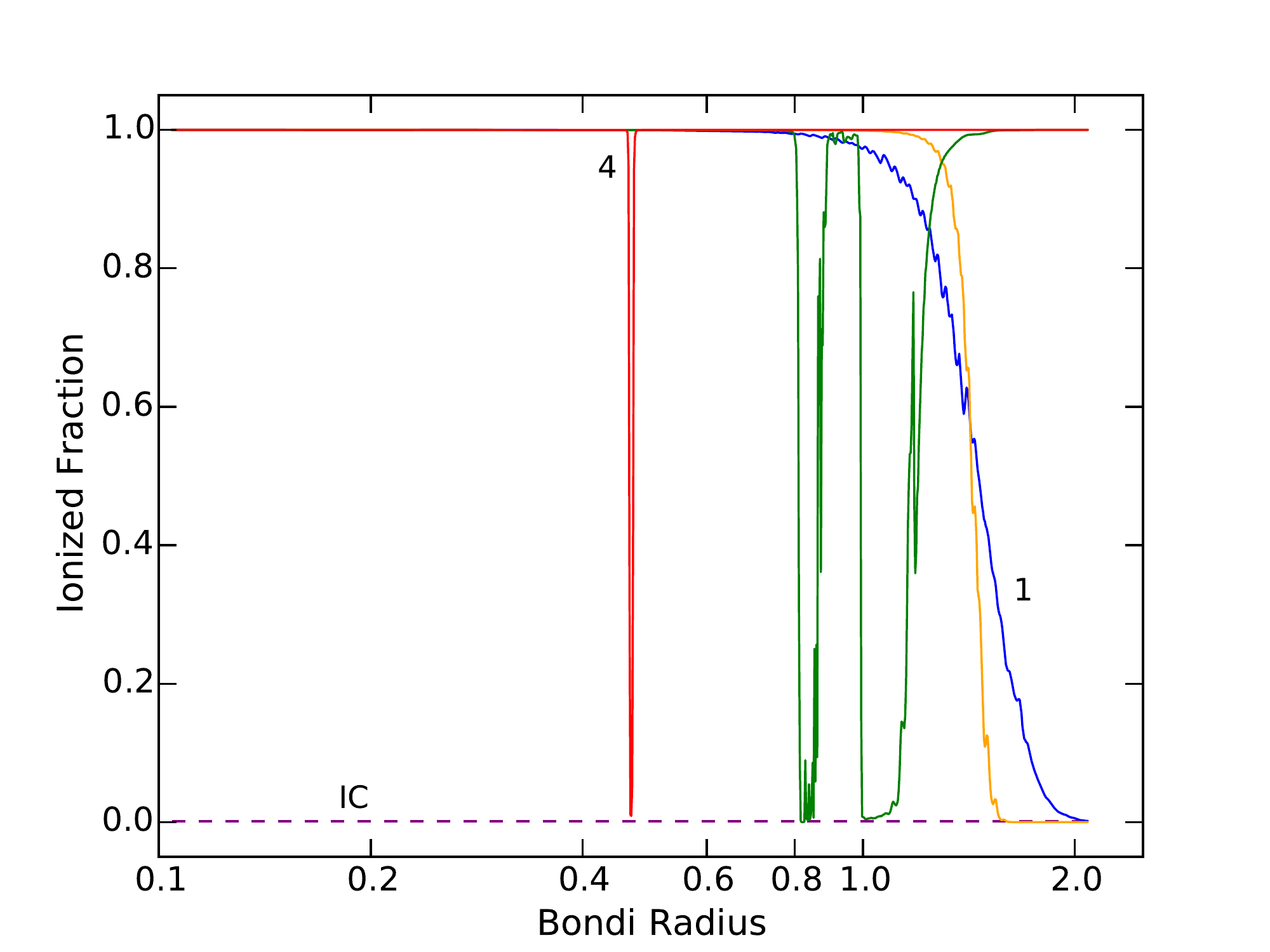}\\
	\vspace{-0.1cm}
	\includegraphics[angle=0,width=0.45\textwidth]{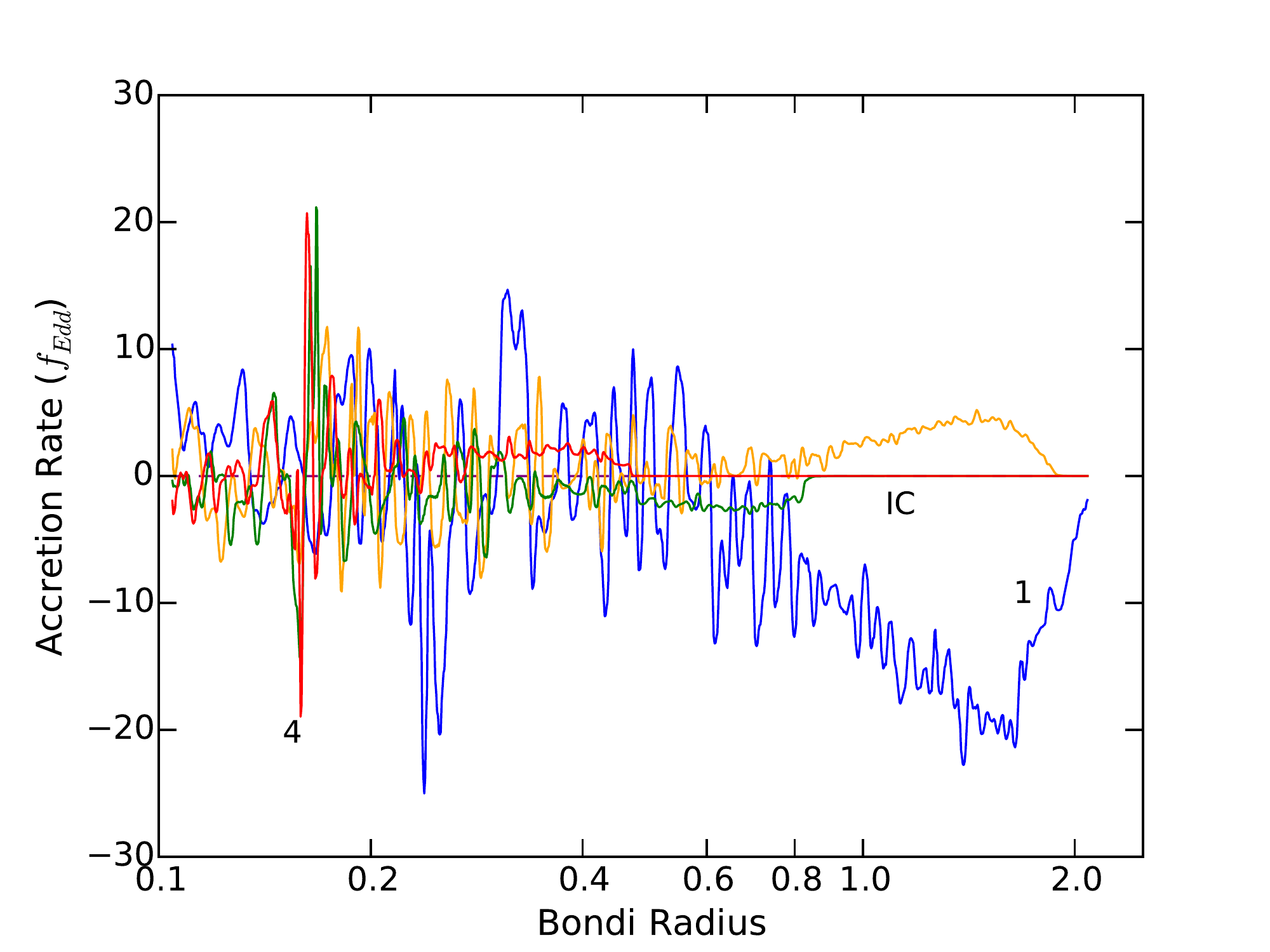}
	\includegraphics[angle=0,width=0.45\textwidth]{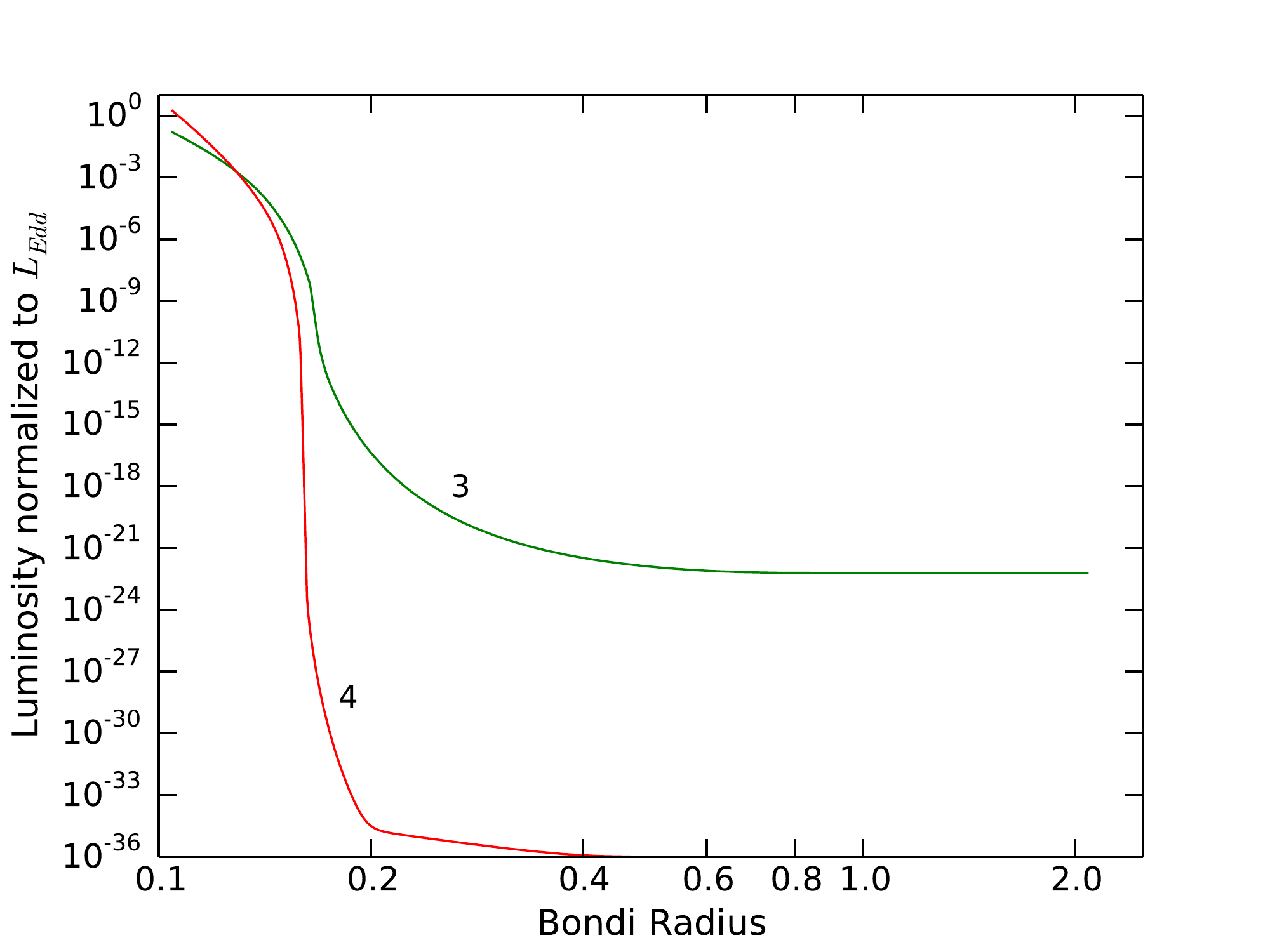}\\
	\vspace{-0.1cm}
	\includegraphics[angle=0,width=0.45\textwidth]{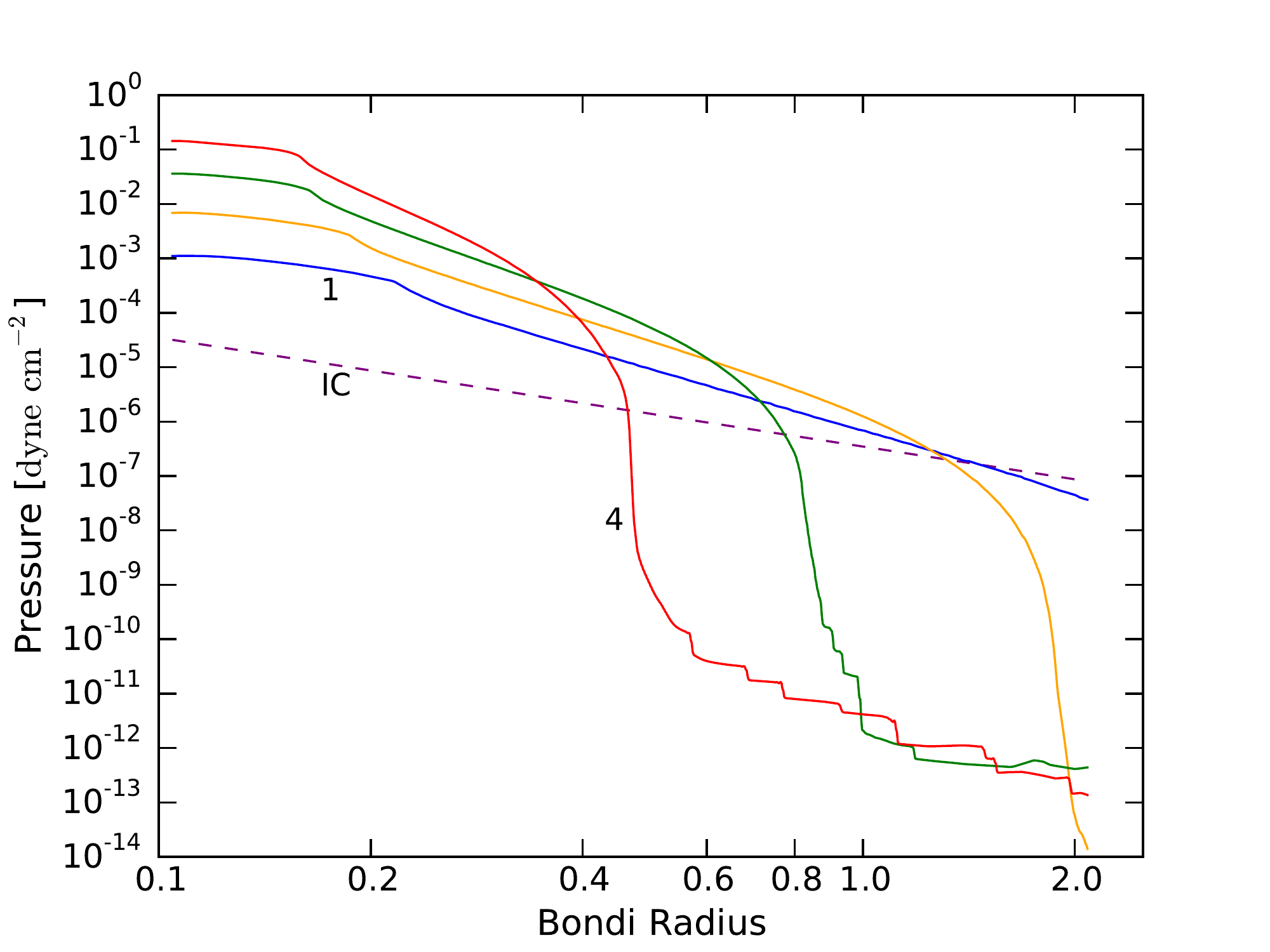}
	\includegraphics[angle=0,width=0.45\textwidth]{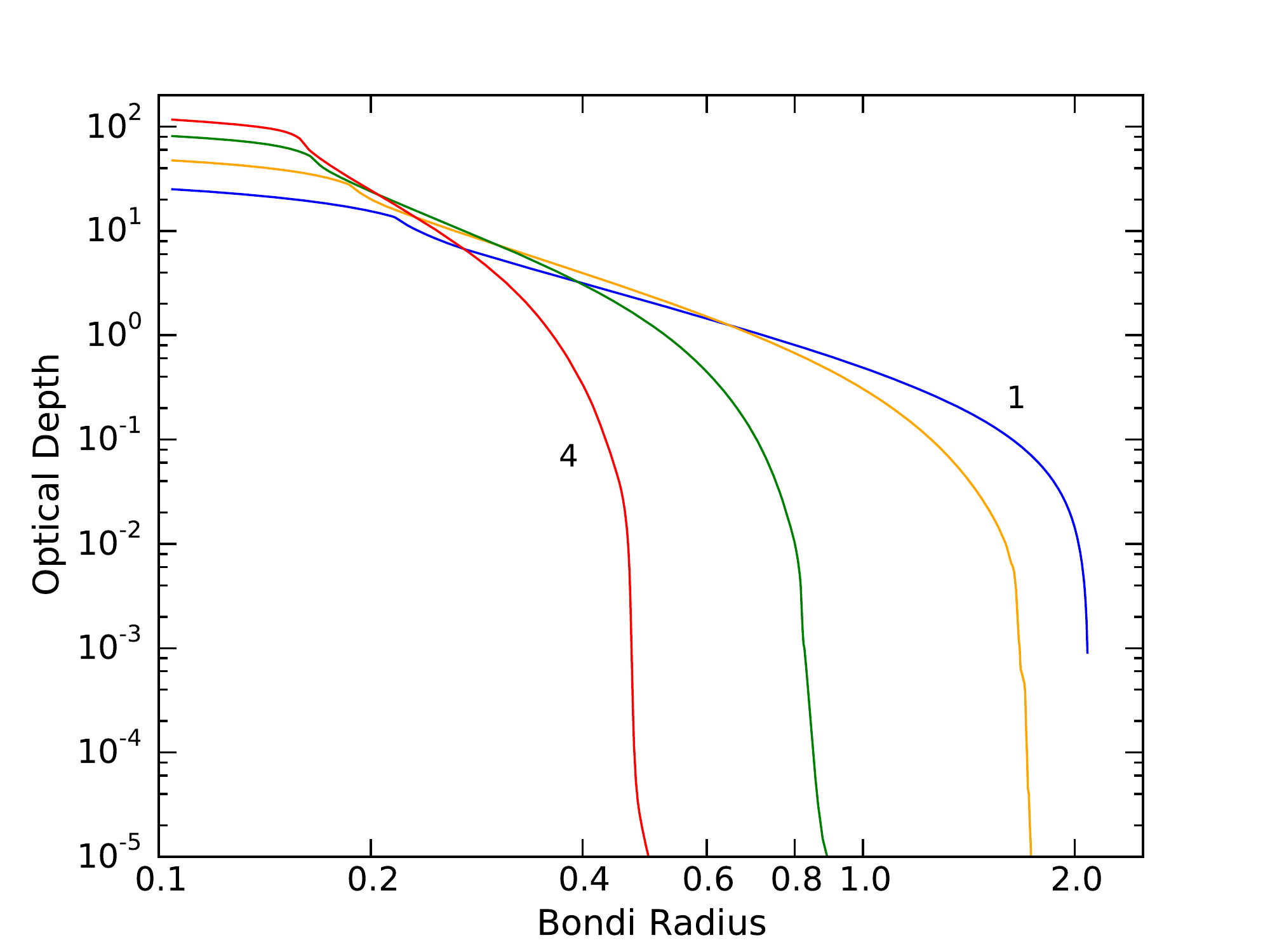}\\	
	\caption{Spatial profiles for the FS simulation: total integration time is $T_{tot} \approx 1.4 \times 10^{8}\, \mathrm{yr}$ and $T_{tot} = i \, \Delta t$ with $\Delta t = 35 \, \mathrm{Myr}$ and $i=1, \, ... \, 4$ (only the labels $i=1$ and $i=4$ are shown). The luminosity lines are present only at data dumps when the IMBH is emitting.}
	\label{fig:FULL_spatial}
\end{figure*}

Fig. \ref{fig:FULL_spatial} shows the spatial profiles for the FS simulation.
From their analysis, three main features are evident: (i) a density wave approaching the center, (ii) strong waves moving towards larger radii, visible from the velocity profile and (iii) the progressive emptying of the outer regions (discussed below).

The density wave, with over-densities as high as $\sim 1$ order of magnitude with respect to the surrounding volume, moves in the inward direction, contrarily to the T2 simulation. This difference is due to the fact that in the FS simulation the radiation pressure and the accretion rate are interconnected, while in the T2 run the former is fixed: this joint evolution leads to a smooth increase in the emitted luminosity and to a different response of the system.
When the IMBH accretes at super-critical rates, the emitted radiation promptly interrupts the mass inflow and consequently the radiation pressure. The accelerated gas moves in the outward direction creating a pressure jump of a factor $\sim 5$ between the two sides of the discontinuity front (see the pressure spatial profile in Fig. \ref{fig:FULL_spatial}). Eventually, the gravitational acceleration of the IMBH inverts the velocity and the accretion starts again.

The radiation pressure affects only a small volume, in the inner section of the accretion flow where $\tau \sim 10-100$, as the optical depth spatial profile shows: the internal layers ($r \ll r_{\tau}$) of the gas distribution are intermittently reached by the radiation pressure, while the external layers ($r \gg r_{\tau}$) are in a quasi free-fall state.
For this reason, similarly to the mechanism detailed in Fig. \ref{fig:mechanism}, the density wave progressively moves towards the center, leading to an increase of $\sim 1$ order of magnitude in the density measured at the innermost cell, as Fig. \ref{fig:Density_velocity_trend} shows.
The top of the density wave $R_{dw}$ moves inward with the time scaling: $R_{dw} \propto t^{-0.7}$.
The optical depth of the inner regions is increased along with the density: this additional effect progressively decreases the volume where the radiation pressure is effective. The luminosity panel is described in a separate subsection below.

The velocity spatial profile shows that the outer regions ($R \gsim 0.5 R_B$) are swept by waves of high-speed ($10-20 \, \mathrm{km \, s}^{-1}$) gas. Note that in some regions of the spatial domain, where the temperature drops near $\sim 10^4 \, \mathrm{K}$, weak shocks are produced in the flow, with Mach numbers ${\cal M} \sim 1.3-2.0$.
This volume, while not affected by radiation pressure, is strongly affected by the thermal pressure exerted from the internal layers: the pressure spatial profile shows, in the external regions, pressure jumps as high as $\sim 6-7$ orders of magnitude. As the radiation pressure is intermittent, the net result is formation of waves in the surrounding gas, whose magnitude and frequency increase with time: an always increasing energy is transported outward with this mechanism.

Finally, the temperature spatial profile shows values as high as $\sim 10^7 \, \mathrm{K}$ in the proximity of the inner boundary, at late stages of accretion, caused by the very large pressure. The temperature spatial profile is reflected in the ionized fraction: the ionized volume expands outward with time.

The complicated behavior of the mass flux spatial profile is a symptom of chaotic motions occurring in the environment, caused mainly by the intermittent irradiation. The last data dump of the FS simulation shows a very small mass flux in the external layers, due to the fact that $\dot{M}_{\bullet}(r) \propto \rho (r)$.

\begin{figure}
\vspace{-1\baselineskip}
\hspace{-0.5cm}
\begin{center}
\includegraphics[angle=0,width=0.5\textwidth]{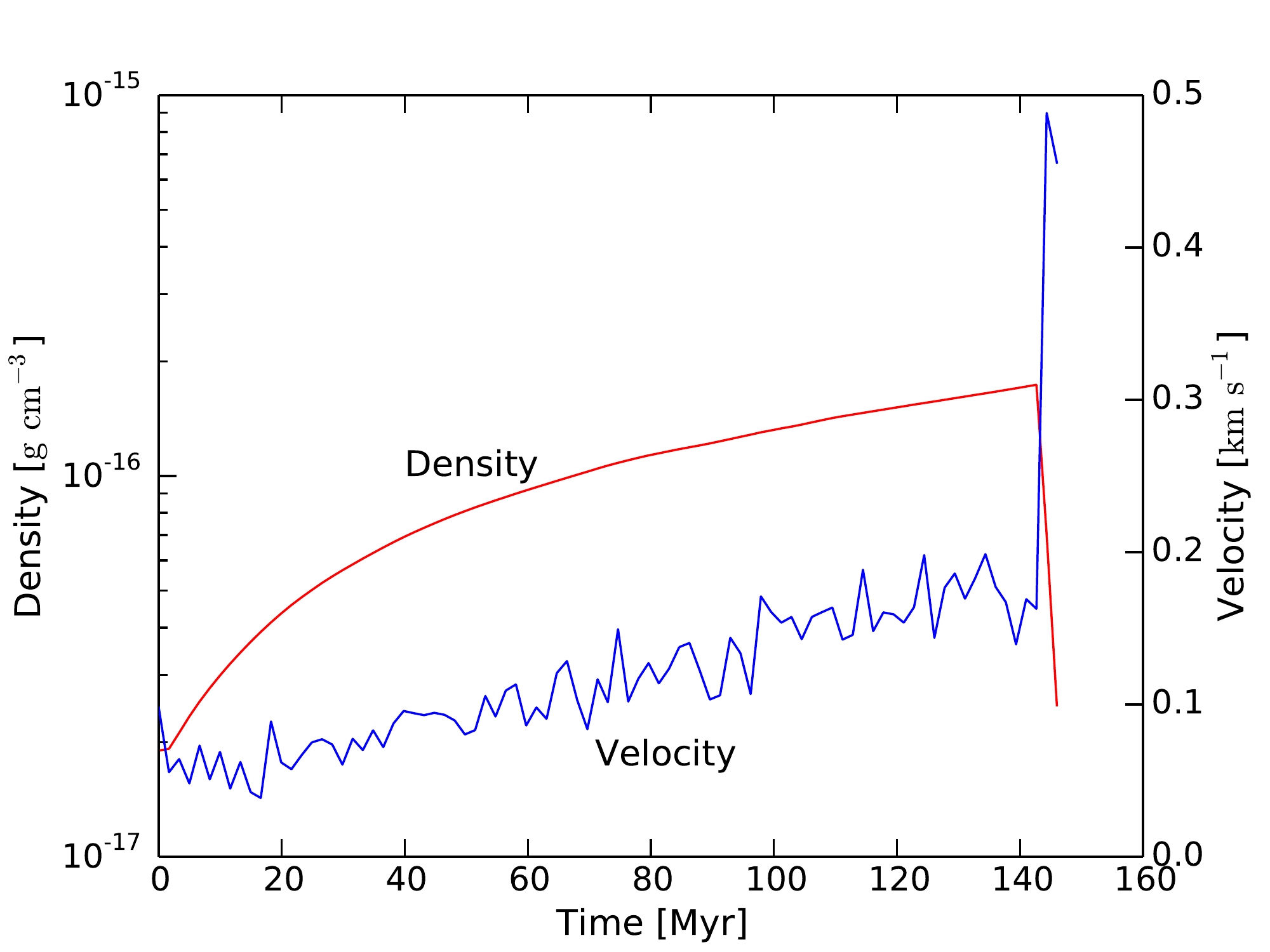}
\caption{Time evolution of density and velocity computed at the innermost cell of the spatial grid. It most clearly shows, along with the following Fig. \ref{fig:Luminosity_accretion_trend}, the final evolutionary phase. At the time $\sim 142 \, \mathrm{Myr}$, a jump of a factor $\sim 3$ in the velocity marks the final act in the evolution of the system: the remaining gas is eventually ejected outward.}
\label{fig:Density_velocity_trend}
\end{center}
\end{figure}

\item{\textbf{Final Burst}: this events marks the end of the accretion flow onto the IMBH, $\sim 142 \, \mathrm{Myr}$ in the simulation. The gas is swept away and the accretion rate goes to zero: the black hole becomes a dark relic, having exhaled its ``last gasp".

As the central density increases, the emitted luminosity rises as well, as shown in Fig. \ref{fig:Luminosity_accretion_trend}, standing always a factor $\sim 2-3$ higher than the Eddington luminosity. This trend is highlighted by the shaded region in the same plot, which shows the value of $f_{Edd}$: after an initial transient period lasting $\sim 20 \, \mathrm{Myr}$ (caused by the necessity to stabilize the accretion flow), the gap with respect to the Eddington luminosity increases with time: the emission is progressively more super-critical. The value of $f_{Edd}$ here is computed including the idle phases: being the duty-cycle $\sim 0.48$, the value of $f_{Edd}$ computed considering only the active phases would be doubled.

The radiated luminosity reaches the value $\sim 3\times 10^{45} \, \mathrm{erg \, s^{-1}}$ ($f_{Edd} \equiv \dot{M}_{\bullet}/\dot{M}_{Edd} = L/L_{Edd} \sim 3$) and the density in the external layers drops: the remaining mass inside the computational domain is a factor $\sim 25$ lower than the initial one, due to both the accretion onto the compact object and the outflow (see below). Differently with respect to the T2 simulation, when the radiation pressure was fixed to a super-critical value, its interconnection with the mass inflow progressively voids the halo outside-in. The external layers are not able to exert a sufficient pressure to contain the expansion of the radiation-driven shell and, as a consequence, the remaining gas is ejected from the system. 

This effect is most clearly visible in Fig. \ref{fig:Density_velocity_trend}, which shows the velocity evolution measured at the innermost cell. The general increasing trend ($\sim 10^{-3} \, {\rm km \, s^{-1} \, Myr^{-1}}$) of the central velocity is abruptly changed by a jump of a factor $\sim 3$. The internal layer starts to move outward with velocities of $\sim 0.5 \, \mathrm{km/s}$.
The same effect is hinted at in Fig. \ref{fig:Luminosity_accretion_trend}, where the accretion (and consequently the emitted luminosity) is abruptly interrupted. After a transient period, the velocity should be re-inverted, but the very low value of the gas mass still inside $R_B$ strongly suggests that the evolution time scale of this system with $M_0 = 10^5 \, \mathrm{\Msun}$ is indeed of order $150 \, \mathrm{Myr}$.}

\begin{figure}
\vspace{-1\baselineskip}
\hspace{-0.5cm}
\begin{center}
\includegraphics[angle=0,width=0.48\textwidth]{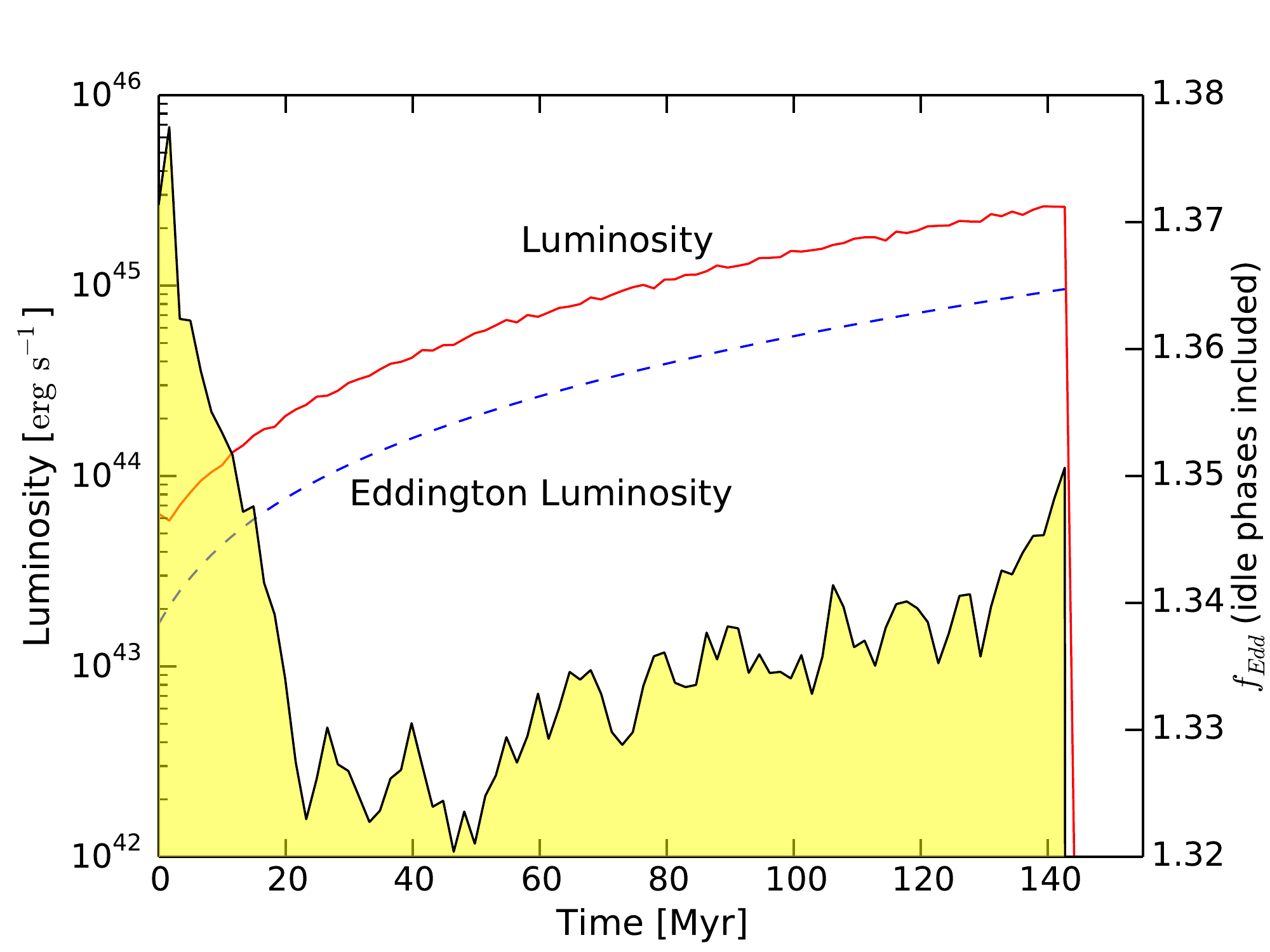}
\caption{Time evolution of the emitted luminosity and of $f_{Edd}$ computed at the innermost cell. The smoothness of the lines is due to an averaging process over a time much longer than the typical idle periods of the accretion. The value of $f_{Edd}$ is computed as a running average over a window period much longer than the typical duration of the duty-cycle and includes the idle phases (while the average values of the luminosity do not). The blue dashed line shows the corresponding time evolution of the Eddington luminosity, which increases as the black hole mass grows. At the time $\sim 142 \, \mathrm{Myr}$ the accretion is abruptly terminated by the final burst.}
\label{fig:Luminosity_accretion_trend}
\end{center}
\end{figure}

}
\end{enumerate}

\subsubsection{Black hole growth}
After having investigated the space and time evolution of the accretion flow, we devote some further analysis to the black hole growth, more specifically to the accretion time scale and to the final mass balance.

An important diagnostic quantity for the accretion process is the duty-cycle, defined in Sec. \ref{sec:methods}. A direct comparison between the three simulations shows that the evolutionary time scales are quite different. As a simple estimator, Table \ref{tab:time_doubles} reports for each simulation the time $t_{2M_0}$ needed for the black hole to double its initial mass, along with the average accretion rate $<\dot{M}>$.
For instance, it is instructive to compare this time scale for the T1 simulation ($\sim 10^4 \, \mathrm{yr}$) and for the FS simulation ($\sim 5 \times 10^6 \, \mathrm{yr}$): the average duty-cycles of these simulations are different, i.e. the FS system spends a smaller fraction of time accreting.
The duty-cycle is strictly dependent on the magnitude of the radiative force, as we have shown in the subsection dedicated to the T2 simulation with simple analytic arguments.
In the FS simulation the duty-cycle stabilizes to an average value $\sim 0.48$, in agreement with the prediction of the approximated Eq. \ref{DC_estimate} (see also \citealt{Ricotti_2012_DC}).

\begin{table*}
\begin{minipage}{170mm}
\begin{center}
\caption{Diagnostic quantities for the two test (T1 and T2) simulations and for the full one (FS), specifically the time $t_{2M_0}$ needed to double the initial mass of the black hole and the average accretion rate $<\dot{M}>$. The final mass for the FS simulation is $\sim 7 \times 10^6 \, \mathrm{\Msun}$.}
\label{tab:time_doubles}
\begin{tabular}{|c|c|c|c|}
\hline\hline
Parameter & T1 & T2 & FS\\
\hline
$t_{2M_0}$ [yr] & $\sim 3\times 10^4$    & $\sim 4 \times10^4$  & $\sim 5\times10^6$  \\
$<\dot{M}>$ [$\mathrm{M_{\odot} \, yr^{-1}}$] & $\sim 3.1$ & $\sim 1.3$ & $\sim 0.1$ \\
\end{tabular}
\end{center}
\end{minipage}
\end{table*}

Furthermore, we have calculated the quantity of gas accreted by the IMBH and the amount ejected from the system, before the final burst.
The density spatial profile in Fig. \ref{fig:FULL_spatial} shows that during the late stages of the simulation the outer layers of the volume are almost empty: in about $\sim 120 \, \mathrm{Myr}$ (the time of the last complete data dump) the density drops by $\sim 7$ orders of magnitude. 
The matter is partly accreted by the central object, partly ejected from the outer boundary of the system by high-speed waves, as described above.
Our final results for the mass balance are summarized in Fig. \ref{fig:mass_trend}. The baryonic mass of the halo is reduced by a factor $\sim 25$ from the beginning of the simulation. Most of this mass ($\sim 90\%$) is accreted onto the black hole, in spite of an average super-Eddington emission, while $\sim 10\%$ is ejected by outflows. Starting from a DCBH of mass $M_0 = 10^5 \, \mathrm{M_{\odot}}$,  the final mass of the black hole is $M_{\bullet} \sim 7 \times 10^6 \, \mathrm{M}_{\odot}$, a fully fledged SMBH.

\begin{figure}
\vspace{-1\baselineskip}
\hspace{-0.5cm}
\begin{center}
\includegraphics[angle=0,width=0.5\textwidth]{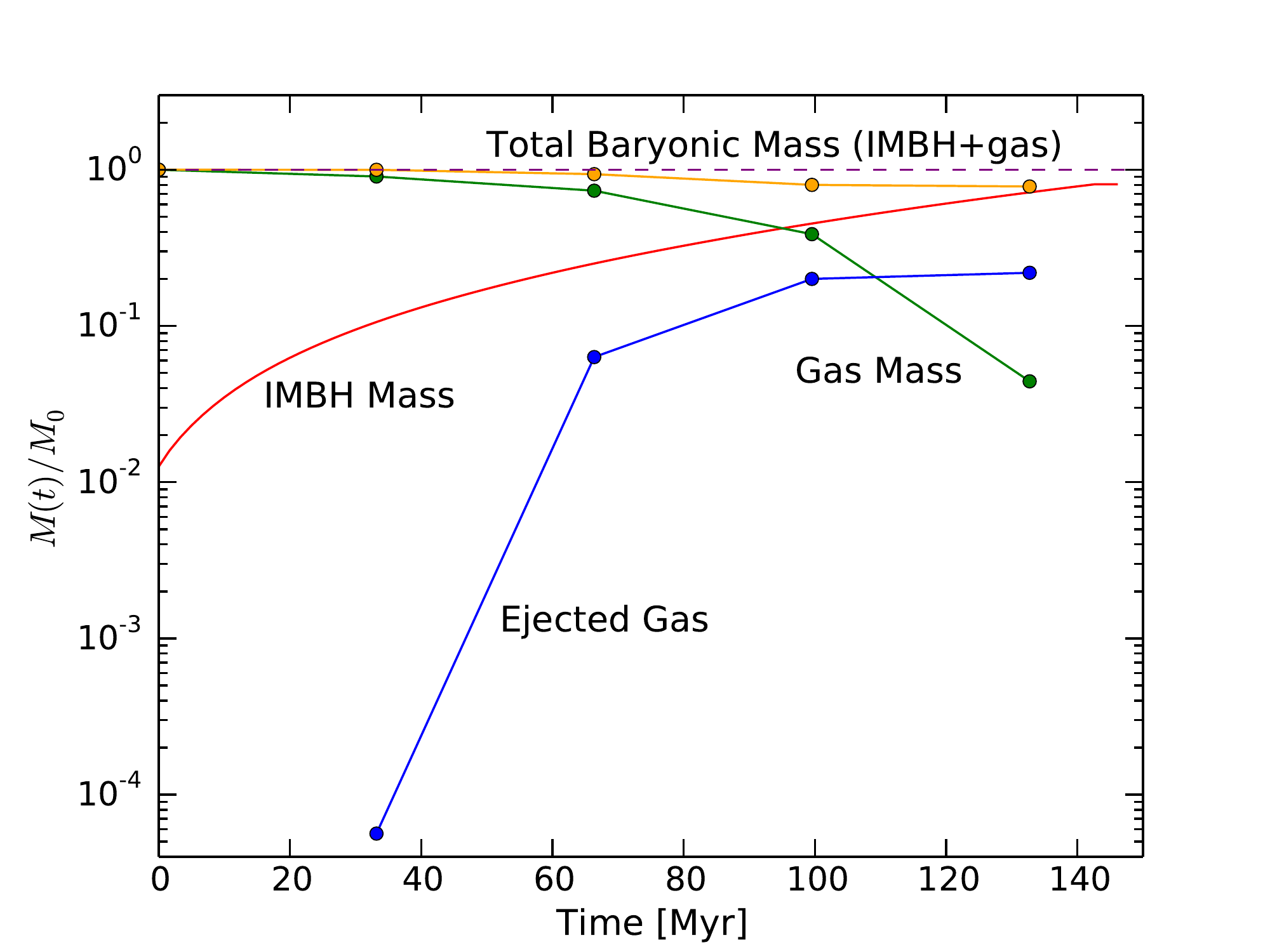}
\caption{Final mass balance for the FS simulation. All masses are normalized to the initial value of the gas mass, $M_{gas} \sim 9.6 \times 10^6 \, \mathrm{M_{\odot}}$. The dashed purple line is only a reference for the unitary value. The IMBH mass line is smoother and extended to a longer time than the others because the corresponding output value is saved at a very high frequency, while other quantities are more discretized in time.}
\label{fig:mass_trend}
\end{center}
\end{figure}

\subsubsection{Radiation emission}
The study of the luminosity emitted from the IMBH is one of the main objectives of this work.
Due to the high values of the optical depth (see its spatial profile in Fig. \ref{fig:FULL_spatial}), the luminosity is almost completely obscured for an external observer. The column density reaches a value of $1.3 \times 10^{25} \, \mathrm{cm^{-2}}$ in the late stages of the FS simulation, as can be roughly estimated from the line number 4 in the density panel of Fig. \ref{fig:FULL_spatial}, considering a mean value $10^{-17} \, \mathrm{g \, cm^{-3}}$ for the mass density ($10^{7} \, \mathrm{cm^{-3}}$ for the number density) between $0.1 R_B$ and $0.5 R_B$.

The spatial profile of the emitted luminosity is shown in Fig. \ref{fig:FULL_spatial} for the two data dumps when the black hole is irradiating.
The luminosity emitted via bremsstrahlung and atomic cooling is, in this spatial range, completely negligible.
The lines show two important facts: (i) the luminosity emitted near the accretion boundary slowly increases, due to the mechanism already detailed in this Section and (ii) the radiation which escapes from the outer boundary decreases and is blocked at progressively smaller radii.
The latter effect is due to the accumulation of matter at lower radial distances, having in mind that $dL/dr \propto - \rho \kappa L$, see Eq. \ref{N3} in the Appendix. In fact, the density measured at the innermost cell increases by a factor $\sim 10$ during the time evolution of the system, as Fig. \ref{fig:Density_velocity_trend} demonstrates. To conclude, the luminosity emitted during the accretion process onto an IMBH at high-z is obscured for most of the system's evolution. We predict that it might be observable during the final burst of radiation.

In order to have a rough estimate of the observability of the latter phase, we suppose that the peak luminosity $L_{peak} \sim 3 \times 10^{45} \, \mathrm{erg \, s^{-1}}$ is emitted in a small ($\Delta \lambda / \lambda \ll 1$) range of Far-IR wavelengths centered at $\lambda = 2 \, \mathrm{\mu m}$ with a flat spectrum (for a detailed study of the contribution of DCBHs sources to the cosmic infrared background, see \citealt{Yue_2013}). From the present study it is hard to estimate the duration of the final burst, due to the lack of the necessary time resolution; however, we can put a solid lower limit of about $5 \, \mathrm{yr}$. This aspect needs to be investigated more thoroughly by future work.
For a source located at $z=10$, the radiation intensity would be ${\cal I} \approx 10^{-6} \, \mathrm{Jy}$, which is observable by the future JWST with only $\sim 100 \, \mathrm{s}$ of integration, yielding a Signal-to-Noise ratio\footnote{Estimate performed with the JWST prototype Exposure Time Calculator (ETC): http://jwstetc.stsci.edu/etc.} of $\sim 250$.

Of course, in order to produce more accurate predictions for the observability, it is necessary to take into account the frequency of these events at high-z and the exact emission spectrum of the source: we defer this task to future work.

\section{Discussion and Conclusions}
\label{sec:disc_concl}
In this work we have investigated the radiation-hydrodynamic evolution of the spherical accretion flow onto a DCBH with initial mass $M_0=10^5 \, \mathrm{M_{\odot}}$ and gas mass in the parent halo $M_{gas} = 9.6 \times 10^6 \, \mathrm{M_{\odot}}$.

The IMBH accretes for $\sim 142 \, \mathrm{Myr}$ with an average duty-cycle $\sim 0.48$ and on average super-critical accretion rates, with $f_{Edd} \simeq 1.35$, i.e. $\dot{M}_{\bullet} \simeq 0.1 \, \mathrm{\Msun \, yr^{-1}}$. The emitted luminosity increases with time, as a consequence of the progressive rise in the mass inflow. The radiation pressure creates strong waves moving, with velocities as high as $\sim 20 \, \mathrm{km \, s^{-1}}$, in the outer ($r \gsim 0.5 \, R_B$) section of the inflow. These waves produce shocks in some regions of the flow.

At the end of the main evolutionary phase $\sim 90\%$ of the gas mass has been accreted onto the compact object, in spite of an average super-Eddington emission, while $\sim 10\%$ has been ejected with outflows. The accretion is terminated when the emitted luminosity reaches the value $\sim 3\times 10^{45} \, \mathrm{erg \, s^{-1}} \sim 3 \, L_{Edd}$ and the related radiation pressure ejects all the remaining gas mass (which, at the final time, is a factor of $\sim 25$ lower than the initial one). We estimate that this final burst of radiation, lasting at least $5 \, \mathrm{yr}$ in the rest-frame, should be observable by the future JWST.
We have identified three different phases of the accretion (the initial phase, the main accretion phase and the final burst), detailing their main characteristics in turn.

We predict that the accretion flow, on average, occurs at mildly super-critical rates for the total evolution of the system (except the very initial transient phase).
Recently, \cite{Alexander_2014}, \cite{Volonteri_2014} and \cite{Madau_2014}, but see also \cite{Volonteri_2005}, have suggested that brief but strongly super-critical accretion episodes (with rates as large as $f_{Edd} \gsim 50$) might explain the rapid black hole mass build-up at $20\gsim z \gsim 7$. Very large and prolonged accretion rates may be sustainable in the so-called ``slim disk" solution (\citealt{Begelman_1982,Paczynski_1982,Mineshige_2000,Sadowski_2009, Sadowski_2011}), an energy-advecting, optically thick flow that generalizes the standard thin disk model \citep{Shakura_Sunyaev_1976}. In these radiatively inefficient models the radiation is trapped (see also the diffusion condition in Appendix) into the high-density gas and is advected inward by the accretion flow: this happens when the photon diffusion time exceeds the time scale for accretion. This would allow a very mild dependence of the emitted luminosity $L$ on the normalized accretion rate $f_{Edd}$, which is usually described as a logarithmic relation (\citealt{Mineshige_2000, Wang_2003}): $L/L_{Edd} \sim 2[1+\ln(f_{Edd}/50)]$, valid for $f_{Edd} \geq 50$.

In our work, we analyze the accretion flow on very large scales ($R \sim R_B$, i.e. the accretion disk is beyond our resolution limit) and photons are never trapped (see the Appendix). For this reason, the accretion flow is radiatively efficient ($L/L_{Edd} \propto f_{Edd}$) and the accretion rates are only mildly super-critical: $f_{Edd} \simeq 1.35$ on average. In our case, the idle phases are caused by the necessity to re-establish the downward accretion flow after the radiation burst, while in the strongly super-critical models they are caused by the need for replenishing the gas reservoirs (e.g. by galaxy mergers, see e.g. \citealt{Volonteri_2014}).
In our simulation, at smaller radial distances ($R \ll R_B$) we expect that the radiation eventually reaches the trapping condition. This is neglected in the present implementation of the simulation, but could critically modify the radiative properties of the source, especially in the light of the aforementioned recent studies.
Some aspects of the simulation may be improved, namely:
\begin{enumerate}
\item The accreting gas, at smaller radii, should form an accretion disk. If the accretion flow has a non-zero angular momentum with respect to the central body, the gas will reach a centrifugal barrier (caused by the steeper radial scaling of the centrifugal acceleration, $\sim r^{-3}$, with respect to the gravitational acceleration, $\sim r^{-2}$) from which it can accrete further inward only if its angular momentum is transported away. This would at least partly modify the irradiation mechanism.
\item A full spectral analysis of the source needs a more accurate description of the interaction between radiation and matter.
\item At smaller radial distances the photons should be trapped and the accretion flow should become energy-advective, i.e. radiatively inefficient, as described above. An appropriate modeling of the inefficient accretion flow would then be required.
\item The magnetic field may also significantly affect the accretion flow structure and behavior, as already pointed out in e.g. \cite{Sadowski_2014} and \cite{McKinney_2014}. The inclusion of an appropriate modeling of the magnetized plasma would then be required.
\end{enumerate}

This work is the basis upon which a full spectral analysis of these sources is to be constructed: this would be the key to unveil the eventual existence of IMBHs during the Cosmic Dawn era.
The existence of IMBHs at high redshifts, although not yet confirmed by observations, would represent a breakthrough in our knowledge of the primordial Universe. Their presence would be also relevant for at least two reasons.
First, the formation of IMBHs in the early Universe would ease the problem of the presence of SMBHs with masses $M_{\bullet} \sim 10^9 \, \mathrm{M_{\odot}}$ at redshifts as high as $z\approx 7.085$ \citep{Mortlock_2011}. These massive seeds would play a role of paramount importance in giving a jump start to the accretion process. Second, the formation of IMBHs at high redshifts could provide a possible interpretation of the near-infrared cosmic
background fluctuations \citep{Yue_2013} and its recently detected cross-correlation with the X-ray background \citep{Cappelluti_2013}. This interpretation would be even more plausible if the primordial population of IMBHs is proved to be highly obscured. The observation of IMBHs could then provide the missing pieces for the solution of these intriguing puzzles.

\vspace{+0.5cm}
We thank L. Ciotti, G. S. Novak and M. Volonteri for useful comments and suggestions.

\section{Appendix}
\label{sec:appendix}
This Appendix contains some more technical details about the modules HD and RT, such as the boundary conditions for velocity and density, the two-stream approximation, the heating and cooling terms and the photon diffusion condition.

\subsection{Additional Boundary Conditions}
In addition to the boundary conditions for the luminosity, already detailed in Sec. \ref{sec:methods}, we need to specify the behavior of velocity and density at the innermost and outermost cells of the spatial grid.
The spatial boundary conditions are: (i) outflow for the inner boundary (with restrictions on velocities, see below) and (ii) void for the outer boundary.
An outflow boundary condition forces the derivatives of the quantities of interest to be zero, i.e. artificially extends the spatial domain outside the boundary.
The restriction for the boundary velocity $v_{b}$ is the following:
\begin{equation}
 v_{b} =
  \begin{cases}
   v(r_{min}) & \text{if  } v(r_{min})<0 \\
   0 & \text{if  } v(r_{min})>0 
  \end{cases}
\end{equation}
and is meant to prevent the replenishment of the computational domain by the gas coming from unresolved spatial scales.
A void boundary condition, on the contrary, constrains the quantity of interest to be zero outside the computational domain: the system composed by the IMBH and its parent halo (up to $\sim 2 R_B$) is isolated in space.

\subsection{The two-stream approximation}
The RT method we used is based on \cite{Novak_2012} and relies on the two-stream approximation, i.e. the luminosity is expressed as the sum of an ingoing and an outgoing radiation stream.
When the optical depth is low, photons of the ingoing stream at any given radius $r_0$ are likely to successfully traverse the inner parts of the halo ($r<r_0$) and emerge as outgoing photons at the same radius, but with $\varphi = \varphi + \pi$ if $\varphi$ is the azimuthal angle. In this case, all the radiation emitted by a source term is to be added to the outgoing stream.
If, instead, the optical depth is large, the ingoing photons are likely to be absorbed for $r<r_0$. Then, only half of the emitted photons should be added to the outgoing stream. The other half should be added to the ingoing stream, where they will in due course be absorbed.
The resulting equations for the two radiation streams are:
\begin{equation}
\frac{dL_{out}}{dr} = 4 \pi r^2 \psi \dot{E} -  \rho \kappa L_{out}
\label{N3}
\end{equation}
\begin{equation}
\frac{dL_{in}}{dr} = 4 \pi r^2(1- \psi) \dot{E} -  \rho \kappa L_{in}
\label{N4}
\end{equation}
where $\psi(r)$ is the fraction of photons emitted at a given radius that are likely to belong to the outgoing stream.
The simplest estimate of the quantity $\psi(r)$ is given in \cite{Novak_2012}:
\begin{equation}
\psi(r) = 1 - \frac{1}{2} \left[\frac{1}{1+e^{-\tau}}\right]\left[\frac{r_1^2}{\mathrm{max}(r_1^2,r^2)}\right]
\end{equation}
where $\tau$ is the optical depth from $r$ to infinity and $r_1$ is the radial distance from the center where $\tau$ reaches unity.
These equations must be complemented with the boundary conditions:
\begin{equation}
  \begin{cases}
  L_{in}(r_{max}) = 0 \\
  L_{out}(r_{min}) = L_{\bullet}
  \end{cases}
\end{equation}
and the expression for the radiative acceleration, Eq. \ref{a_rad}, must be changed into the following one:
\begin{equation}
a_{rad} = \frac{\kappa(\rho, T) (L_{out} - L_{in})}{4 \pi r^2 c}
\end{equation}

\subsection{Heating and Cooling}
This paragraph deals with the non-adiabatic regime, i.e. the source/sink term $\dot{E}$ (Eqs. \ref{N1}-\ref{N2} and \ref{N3}-\ref{N4}). Differently from the implementation in \cite{Novak_2012}, where $\dot{E}$ accounts for the energy transfer among different frequency bands, in our case the interaction between matter and radiation is purely elastic (i.e. the frequency of the interacting photon is unchanged) and this term expresses the energy emitted or absorbed by matter per unit time and per unit volume.

For a gas at $T \lesssim 10^4\, \mathrm{K}$ (i.e. below the atomic hydrogen cooling threshold) the energy equation is purely adiabatic.
Instead, for a gas at $T \gsim 10^4\, \mathrm{K}$, the term $\dot{E}$ takes into account the bremsstrahlung cooling and the atomic cooling (recombination and collisional excitation, for H and He):
\begin{equation}
\dot{E} = 2 \left(\psi - \frac{1}{2}\right) \chi_{ion} n^2 {\cal S}
\end{equation}
Here, $n$ is the number density and $\psi$ is the fraction of photons emitted at a given radius that are likely to belong to the outgoing stream (see above in this Appendix). The term $2 \left(\psi - \frac{1}{2}\right)$ is the fraction of transmitted (i.e. not absorbed) photons, since $\psi$ is defined as $\psi = (1+p_{trans}/2)$, where $p_{trans}$ is the transmission probability. The term ${\cal S}$ includes all the coefficients for bremsstrahlung and atomic cooling (in units $\mathrm{erg \, cm^{3} \, s^{-1}}$ as reported in \citealt{Maselli_03} and \citealt{Cen_1992}).

Moreover, the term $\chi_{ion}$ is the ionized fraction, which accounts for the fraction of atoms that can contribute to the bremsstrahlung.
This last term is calculated as $\chi_{ion} = \left(\alpha_b/\gamma + 1\right)^{-1}$, where $\alpha_b$ is the recombination rate and $\gamma$ is the collisional ionization rate. Both of them are expressed in $\mathrm{cm}^3 \mathrm{s}^{-1}$ and are defined as (see e.g. \citealt{Maselli_03}):
\begin{equation}
\alpha_b = \frac{8.4\times 10^{-11}}{\sqrt{T}} \left(\frac{T}{1000}\right)^{-0.2} \left[1.0+\left(\frac{T}{10^6}\right)^{0.7}\right]^{-1}
\end{equation}
\begin{equation}
\gamma = 1.27\times10^{-11} \sqrt{T} \, e^{-157809.1/T} \left[1.0+\left(\frac{T}{10^5}\right)^{0.5}\right]^{-1}
\end{equation} 
 
\subsection{Photon Diffusion Condition} 
Given the very high values of the optical depth (as high as $\sim 100$ for the FS simulation), it is important to check whether a proper treatment for photon diffusion is required.
Photons in the diffusive regime are advected inward with the gas, rather than being diffused out of the accretion flow.
In this condition, the trapped infalling material should be considered a ``quasi-star", similar to the phase advocated by \cite{Begelman_2008}, with an atmosphere in Local Thermodynamic Equilibrium: the emission from the inner section of the accreting flow would then be thermal.

\cite{Begelman_1978} gives a very practical way to assess the occurrence of the diffusive behavior of photons. Photons displaced at some radius $r$ are trapped-in if:
\begin{equation}
\tau(r) \, \frac{v(r)}{c} > 1
\end{equation}
Throughout the paper, we refer to this formula as the diffusion condition, which is never met in our grid, because the maximum values reached are of order $\tau(r) \, v(r)/c \approx 10^{-3}$.
This proves that the photons inside the simulated spherical volume in our work are never in the diffusive regime.
It is likely that this condition is actually met at smaller radii, where the gas may form a structure quite similar to a stellar atmosphere. We defer the investigation of this possibility to future work.


\end{document}